\documentclass[preprint,aps,amsmath,nofootinbib,tightenlines]{revtex4}
\usepackage{graphicx}
\usepackage{bm}
\usepackage{epsfig}

\usepackage[hypertex]{hyperref}

\def\Dslash{D\!\!\!\!\slash}

\def\nslash{n\!\!\!\slash}
\def\bnslash{\bar n\!\!\!\slash}

\def\vslash{v\!\!\!\slash}

\def\OMIT#1{}

\newcommand{\nn}{\nonumber} 

\newcommand{\bn}{{\bar n}}
\newcommand{\bea}{\begin{eqnarray}}
\newcommand{\eea}{\end{eqnarray}}

\newcommand{\bnP}{\bar {\cal P}}

\newcommand{\mcdot}{\!\cdot\!}

\def\babar{\mbox{\sl B\hspace{-0.4em} {\small\sl A}\hspace{-0.37em} \sl 
B\hspace{-0.4em} {\small\sl A\hspace{-0.02em}R}}}

\begin{document}


\preprint{ \vbox{ \hbox{hep-ph/0511334}
 \hbox{MIT-CTP-3691} }}

\title{\phantom{x} \vspace{0.5cm} \boldmath
  Shape-Function Effects
  and Split Matching in $B\to X_s \ell^+\ell^-$
 \vspace{0.6cm} }

\author{Keith S. M. Lee}
\affiliation{Center for Theoretical Physics, Laboratory for Nuclear Science, \\ Massachusetts Institute of
Technology,
Cambridge, MA 02139\footnote{Electronic address: ksml@mit.edu, iains@mit.edu}
\vspace{0.2cm}}
\author{Iain W. Stewart\vspace{0.4cm}}
\affiliation{Center for Theoretical Physics, Laboratory for Nuclear Science, \\ Massachusetts Institute of
Technology,
Cambridge, MA 02139\footnote{Electronic address: ksml@mit.edu, iains@mit.edu}
\vspace{0.2cm}}

\vspace{0.2cm}
\vspace{1cm}

\begin{abstract}
  \vspace{0.3cm} 

  We derive the triply differential spectrum for the inclusive rare decay $B \to
  X_s \ell^+\ell^-$ in the shape function region, in which $X_s$ is jet-like
  with $m_X^2 \lesssim m_b \Lambda_{\rm QCD}$. Experimental cuts make this a
  relevant region.  The perturbative and non-perturbative parts of the matrix
  elements can be defined with the Soft-Collinear Effective Theory, which is
  used to incorporate $\alpha_s$ corrections consistently.  We prove that, with a
  suitable power counting for the dilepton invariant mass, the same universal
  jet and shape functions appear as in $B\to X_s\gamma$ and $B\to
  X_u\ell\bar\nu$ decays.  Parts of the usual $\alpha_s(m_b)$ corrections go
  into the jet function at a lower scale, and parts go into the non-perturbative
  shape function. For $B\to X_s \ell^+\ell^-$, the perturbative series in
  $\alpha_s$ are of a different character above and below $\mu=m_b$. We
  introduce a ``split matching'' method that allows the series in these regions
  to be treated independently.

\end{abstract}

\maketitle

\section{Introduction} \label{sect_intro}

The $B$ meson is particularly suitable for probing QCD and flavor physics in the
Standard Model, since the large mass of the $b$ quark relative to $\Lambda_{\rm
  QCD}$ provides a useful expansion parameter, $\Lambda_{\rm QCD}/m_b\sim 0.1$.
The study of inclusive $B$ decays circumvents the need for precision hadronic
form factors, while still allowing model-independent predictions. Rare inclusive
decays, which involve flavor-changing neutral currents (FCNCs), not only allow
measurements of CKM matrix elements, in particular $V_{ts}$ and $V_{td}$, but
are also highly sensitive to new physics, since they do not occur at tree level
in the Standard Model.

Among the inclusive rare $B$ decays, the radiative process $B \to X_s\gamma$ has
received the most attention, having been measured first by CLEO \cite{cleo95}
and subsequently by other experiments \cite{aleph98,belle01,cleo01,babar02}.
These measurements have provided significant constraints on extensions to the
Standard Model. The decay $B \to X_s \ell^+\ell^-$ is complementary to, and more
complicated than, $B \to X_s\gamma$. Its potential for revealing information
beyond that supplied by the radiative decay is due to the presence of two extra
operators in the effective electroweak Hamiltonian and the availability of
additional kinematical variables, such as the dilepton invariant-mass spectrum
and the forward-backward asymmetry. Belle and {\babar} have already made
initial measurements of this dilepton process \cite{belle03,babar03}.

Provided that one makes suitable phase-space cuts to avoid $c\bar c$ resonances,
$B \to X_s \ell^+\ell^-$ is dominated by the quark-level process, which was
calculated in Ref.~\cite{gsw89}. Owing to the disparate scales, $m_b \ll m_W$,
one encounters large logarithms of the form $\alpha_s^n(m_b)\log^n(m_b/m_W)\;$
(leading log [LL]), $\alpha_s^{n+1}(m_b)\log^n(m_b/m_W)\;$ (next-to-leading log
[NLL]), \ldots, which should be summed. The NLL calculations were completed in
Refs.~\cite{bm95,misiak93}, and the NNLL analysis, although technically not
fully complete, is at a level that the scale uncertainties have been substantially
reduced, after the combined efforts of a number of groups
\cite{bmu00,aagw02,ghiy03,bggh04,ghiy04}.

Non-perturbative corrections to the quark-level result can also be calculated by
means of a local operator product expansion (OPE)~\cite{Shifman}, with
non-perturbative matrix elements defined with the help of the Heavy Quark
Effective Theory (HQET)~\cite{hbook}.  As is the case for $B \to X_s\gamma$ and
$B \to X_u \ell\bar\nu$, there are no ${\cal O}(1/m_b)$ corrections.  The ${\cal
  O}(1/m_b^2)$ corrections and OPE were considered in Ref.~\cite{fls94} and
subsequently corrected in Ref.~\cite{ahhm97}. The ${\cal O}(1/m_b^3)$
corrections were computed in Ref.~\cite{bb00}.  There are also non-perturbative
contributions arising from the $c\bar c$ intermediate states.  The largest
$c\bar c$ resonances, i.e.\ the $J/\psi$ and $\psi'$, can be removed by suitable
cuts in the dilepton mass spectrum.  It is generally believed that the
operator product expansion holds for the computation of the dilepton invariant
mass as long as one avoids the region with the first two narrow resonances,
although no complete proof of this (for the full operator basis) has been given. A
picture for the structure of resonances can be obtained using the model of
Kr\"{u}ger and Sehgal \cite{ks96}, which estimates factorizable contributions
based on a dispersion relation and experimental data on $\sigma(e^+ e^- \to
c\bar c \,+ \mbox{hadrons})$.  Non-factorizable effects have  been estimated
in a model-independent way by means of an expansion in $1/m_c$ \cite{bir98},
which is valid only away from the resonances.

Staying away from the resonance regions in the dilepton mass spectrum leaves two
perturbative windows, the low- and high-$q^2$ regions, corresponding to $q^2 \le
6 \, \mbox{GeV}^2$ and $q^2 \ge 14.4\, \mbox{GeV}^2$ respectively.  These have
complementary advantages and disadvantages \cite{ghiy04}. For example, the
latter has significant $1/m_b$ corrections but negligible scale and charm-mass
dependence, whereas the former has small $1/m_b$ corrections but non-negligible
scale and charm-mass dependence. The low-$q^2$ region has a high rate compared
to the high-$q^2$ region and so experimental spectra will become precise for
this region first. However, at low $q^2$ an additional cut is required, making
measurements less inclusive. In particular, a hadronic invariant-mass cut is
imposed in order to eliminate the combinatorial background, which includes the
semileptonic decay $b \to c (\to s e^+\nu) \, e^- \bar\nu = b \to s e^+ e^- +
\mbox{missing energy}$.  The latest analyses from {\babar} and Belle impose cuts
of $m_X\le 1.8\,{\rm GeV}$ and $m_X\le 2.0\,{\rm GeV}$
respectively~\cite{belle03,babar03,Exptcut}, which in the $B$-meson rest frame 
correspond to
$q^0 \gtrsim 2.3\,{\rm GeV}$ and put the decay rate in the so-called shape
function region~\cite{Neubert:1993ch}. This cut dependence has so far been
analyzed only in the Fermi-motion model~\cite{ah98}.  

Existing calculations for $B\to X_s\ell^+\ell^-$ are based on a local operator
product expansion in $\Lambda_{\rm QCD}/m_b$.  When $m_X^2 \lesssim m_b \Lambda
\sim (2\,{\rm GeV})^2$, this operator product expansion breaks down, and, instead of
depending on non-perturbative parameters ($\lambda_1,\lambda_2, \ldots$) that
are matrix elements of local operators, the decay rates depend on
non-perturbative functions. Furthermore, in this region the standard
perturbative $\alpha_s$ corrections to the partonic process $b\to s\ell^+\ell^-$
do not apply, since some of these corrections become non-perturbative.  Thus,
even at leading order there does not exist in the literature a model-independent
computation of the $B\to X_s\ell^+\ell^-$ decay rate that can be compared
directly with the data at low $q^2$.

Here we study $B \to X_s \ell^+ \ell^-$ ($\ell = e, \mu$) in the shape function
region for the first time. The relevant scales are $m_W^2 \gg m_b^2 \gg
m_b\Lambda_{\rm QCD} \gg \Lambda_{\rm QCD}^2$. In this paper we derive the
proper theoretical expression for the leading-order triply differential decay
rate, which incorporates non-perturbative effects that appear at this order and
a correct treatment of the perturbative corrections at each of the scales.
Using the Soft-Collinear Effective Theory (SCET)~\cite{bfl01,bfps01,bs1,bps02}
we prove that the non-perturbative dynamics governing the measurable low-$q^2$
spectra in $B\to X_s \ell^+ \ell^-$ is determined by the same universal shape
function as in endpoint $B\to X_u\ell\bar\nu$ and $B\to X_s\gamma$ decays.  We
also prove that the decay rate can be split into a product of scale-invariant
terms, capturing physics at scales above and below $m_b$.
We show that this procedure, which we call ``split matching'', can be used to 
deal with a tension between the perturbative corrections that come from these 
two regions. Implications for relating the $B\to X_s\ell^+\ell^-$ measurements 
with the $m_X$ cut to the Wilson coefficients are presented in a companion 
publication~\cite{llst}.

In the shape function region, the set of outgoing hadronic states becomes
jet-like and the relevant degrees of freedom are collinear and ultrasoft modes.
This is why the appropriate theoretical method is SCET.  The endpoint
region has been the focus of much work in the context of $B \to X_s\gamma$ and
$B \to X_u \ell \bar\nu$ (see e.g.
Refs.~\cite{Neubert:1993ch,ks94,bps02,Leibovich:1999xf,Leibovich:2000ig,Mannel,bm04,blnp04,blm1,Leibovich:2002ys,blm2,klis04,Bosch,Beneke,Lange}).
In $B\to X_u\ell\bar\nu$ this is because of the cuts used to eliminate the
dominant $b\to c$ background.  In $B\to X_s\gamma$, it is known that cuts with
$q^0\gtrsim 2.1\,{\rm GeV}$ put us in the shape function region.\footnote{In
  Ref.~\cite{neubert} it was pointed out that even a cut of $E_\gamma \ge E_0 =
  1.8\,{\rm GeV}$, corresponding to $m_X \lesssim 3\,{\rm GeV}$, might not
  guarantee that a theoretical description in terms of the local OPE is
  sufficient, owing to sensitivity to the scale $\Delta = m_b-2E_0$ in power and
  perturbative corrections.  Using a multi-scale OPE with an expansion in
  $\Lambda/\Delta$ allows the shape function and local OPE regions to be
  connected~\cite{bm04,blnp04,neubert}.}

\begin{figure}[t!]
  \vskip .2cm \centerline{ \mbox{\epsfysize=8.truecm
      \hbox{\epsfbox{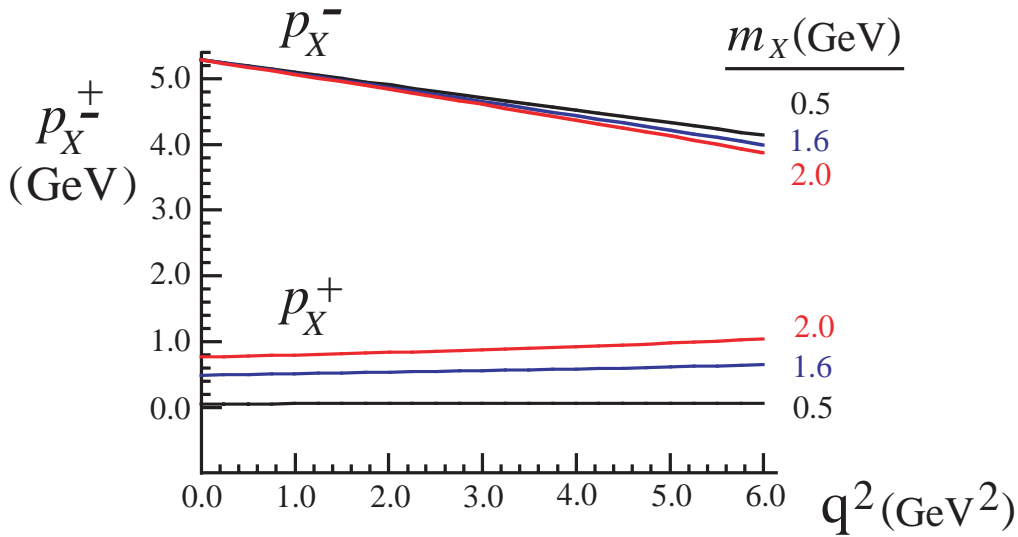}} } } \vskip -.4cm {\caption[1]{The
      kinematic range for $p_X^-$ and $p_X^+$ given the experimental cuts of
      $q^2 < 6\,{\rm GeV}^2$ and $m_X \le 2.0\,{\rm GeV}$ for $B\to
      X_s\ell^+\ell^-$.}
\label{fig_kinematics} }
 \vskip -0.1cm
\end{figure}

In the small-$q^2$ region of $B\to X_s\ell^+\ell^-$ with $q^0 \ge 2.3\,{\rm
  GeV}$, shape-function effects also dominate rather than the expansion in local
operators.  To see this, we note that the $m_X$ cut causes $2 m_B E_X = m_B^2 +
m_X^2 - q^2 \gg m_X^2$.  Decomposing $2E_X= p_X^+ + p_X^-$ with $m_X^2=p_X^-
p_X^+$, we see that the $X_s$ is jet-like with $p_X^- \gg p_X^+$, and the
restricted sum over states in the $X_s$ causes the non-perturbative shape
functions to become important. For the experimental cuts on $q^2$ and $m_X$,
values for $p_X^\pm$ are shown in Fig.~\ref{fig_kinematics}. It should be clear
from this figure that the measurable spectrum is dominated by decays for which
$p_X^- \gg p_X^+$.

To compute $B\to X_s\ell^+\ell^-$ in the shape function region with
renormalization-group evolution requires the following steps:\ 
\begin{itemize}
 \item[i)] matching the Standard Model at $\mu\simeq m_W$ on to $H_W$, 
 \item[ii)] running $H_W$ to $\mu\simeq m_b$, 
\item[iii)] matching at $\mu \simeq m_b$ on to operators in SCET, 
\item[iv)] running in SCET to $\mu\simeq \sqrt{m_b\Lambda}$,
\item[v)] computation of the imaginary part of forward-scattering time-ordered
  products in SCET at $\mu\simeq \sqrt{m_b\Lambda}$. This leads to a separation
  of scales in a factorization theorem, which at LO takes the form\footnote{Note
    that the operator product expansion used here occurs at $\mu\simeq
    \sqrt{m_b\Lambda}$, rather than at $m_b^2$, as in the standard local OPE.}
  $$d^3\Gamma^{(0)} = H \int\! dk\: {\cal J}^{(0)}(k)\: f^{(0)}(k) \,,$$
  with
  perturbative $H$ and ${\cal J}^{(0)}$, and the LO non-perturbative shape 
  function $f^{(0)}$,
\item[vi)] evolution of the shape function $f^{(0)}$ from $\Lambda_{\rm QCD}$ up
  to $\mu\simeq \sqrt{m_b\Lambda_{\rm QCD}}$\,.
\end{itemize}
For the shape-function decay rate, steps i-ii) are the same as the local OPE
results for $B\to X_s\ell^+\ell^-$. Furthermore, based on the structure of
leading-order SCET operators that we find for $B\to X_s\ell^+\ell^-$, we
demonstrate that results for other inclusive endpoint analyses can be used in
steps iv) and vi)~\cite{bfl01,bfps01,blnp04}.\footnote{In step iv) we can run
  the hard functions down using results from Refs.~\cite{bfl01,bfps01}. In step
  vi) we can run the shape function up to the intermediate scale using the
  simple result from Ref.~\cite{blnp04}. An equally valid option would be to
  evolve the perturbative parts of the rate down to a scale $\mu\simeq 1\,{\rm
    GeV}$, as considered earlier~\cite{Leibovich:1999xf,Ira,bm04,bfl01}.}  
Because of this our
computations focussed on steps iii) and v). In step iii) we show how to
implement the split-matching procedure to formulate the perturbative 
corrections, which we elaborate on below. In step v) we derive a factorization 
theorem for $B\to X_s\ell^+\ell^-$. This includes computing the hard 
coefficient functions $H$ at NLL order and formulating the structure of these 
terms to all orders in $\alpha_s$. It also includes a derivation of formulae 
for the decay rate and forward-backward asymmetry that properly take into 
account the effect of the current experimental cuts and the perturbative and 
non-perturbative corrections.

At leading order in the power expansion the result of steps i)-vi) takes the
schematic form
\begin{eqnarray} \label{runrate}
d^3 \Gamma^{(0)} & = & {\cal E}(\mu_W) U_W(\mu_W, \mu_0) 
               {\cal B}(\mu_0) U_H(\mu_0,\mu_i) {\cal J}(\mu_i) 
                               U_S(\mu_i,\mu_\Lambda)f^{(0)}(\mu_\Lambda) 
  \,,\nn\\[5pt]
 && \mu_W \simeq m_W\,,\ \mu_0 \simeq m_b \,,\ 
 \mu_i \simeq (m_b\Lambda)^{1/2}\,,\  \mu_\Lambda \simeq 1\,{\rm GeV} \,,
\end{eqnarray}
where ${\cal E}$, ${\cal B}$ and ${\cal J}$ represent matching at various
scales, and $U_W$, $U_H$ and $U_S$ represent the running between these scales.
Eq.~(\ref{runrate}) shows only the scale dependence explicitly, not the
kinematic dependences or the convolutions between ${\cal J}$, $U_S$, and
$f^{(0)}$, which we describe later on.

In a standard application of renormalization-group-improved perturbation theory
(LL, NLL, NNLL, etc.), the results at each stage of matching and running are
tied together, as depicted in Eq.~(\ref{runrate}).  Usually this would not be a
problem, but for $B\to X_s\ell^+\ell^-$ the nature of the perturbative
expansion above and below $\mu\simeq m_b$ is different.  Above $\mu\simeq m_b$
the series of $(\alpha_s \ln)^k$ terms are of the traditional form, with a basis
of $\sim 10$ operators (including four-quark operators), whose mixing is
crucial.  Below $\mu\simeq m_b$ we demonstrate that the evolution is universal
(to all orders in $\alpha_s$) for the leading-order operators, but there are
Sudakov double logarithms of the ratios of scales, which give a more complicated
series.  It turns out to be convenient to decouple these two stages of
resummation so that one can consider working to different orders in the
$\alpha_s$ expansion above and below $\mu=m_b$. There is a simple reason why
this decoupling is important: for $\mu\ge m_b$ the power counting and running
are for currents in the electroweak Hamiltonian and dictate treating $C_9\sim
1/\alpha_s$ with $C_7\sim 1$ and $C_{10}\sim 1$. However, at $\mu=m_b$ the
coefficients $C_9$ and $C_{10}$ are numerically comparable. For $\mu \le m_b$ in
the shape function region we must organize the power counting and running for
time-ordered products of currents in SCET rather than amplitudes, and it would
be vexing to have to include terms $\propto C_9^{\,2}$ to ${\cal O}(\alpha_s^2)$
before including the $C_{10}^{\,2}$ and $C_7^{\,2}$ terms at order ${\cal
  O}(\alpha_s^0)$.  Thus, once we are below the scale $m_b$, a counting with
$C_9\sim C_{10}\sim C_7\sim 1$ is more appropriate.

To decouple these two regions for $B\to X_s\ell^+\ell^-$ decays we make use of
two facts: i) for $\mu \ge m_b$ the operator ${\cal O}_{10}$ involves a
conserved current and has no operators mixing into it, so it does not have an
anomalous dimension, and ii) for $\mu \le m_b$ all LO biquark operators in the
Soft-Collinear Effective Theory have the same anomalous dimension~\cite{bfps01}.
We shall show that the operators for $B\to X_s\ell^+\ell^-$ are related to these
biquark operators.  These properties ensure that we can separate the
perturbative treatments in these two regions at any order in perturbation
theory. This is done by introducing two matching scales, $\mu_0\simeq m_b$ and
$\mu_b\simeq m_b$.  The two aforementioned facts allow us to write
\begin{align}
  U_W(\mu_W,\mu_0) {\cal B}(\mu_0) U_H(\mu_0,\mu_i) &=
    U_W(\mu_W,\mu_0) {\cal B}(\mu_0,\mu_b) U_H(\mu_b,\mu_i) \nn\\
    & = U_W(\mu_W,\mu_0) B_1(\mu_0) B_2(\mu_b) U_H(\mu_b,\mu_i)   \,,
\end{align}
with well-defined $B_1$ and $B_2$. We define $B_2(\mu_b)$ by using the matching
for the operator ${\cal O}_{10}$ and extend this to find $B_2$ matching
coefficients for the other operators using property ii) above. The remaining
contributions match on to $B_1$. Diagrams which are related to the anomalous
dimension for $\mu\ge m_b$ end up being matched at the scale $\mu_0$ on to
$B_1$, while those related to anomalous dimensions for $\mu \le m_b$ are matched
at a different scale, $\mu_b$, on to $B_2$.  This leaves
\begin{align} \label{mu2}
  d^3 \Gamma^{(0)} & = & \Big[ {\cal E}(\mu_W) U_W(\mu_W, \mu_0)
               B_1(\mu_0) \Big] \Big[ B_2(\mu_b) U_H(\mu_b,\mu_i) {\cal J}(\mu_i)
                               U_S(\mu_i,\mu_\Lambda)f^{(0)}(\mu_\Lambda)
     \Big] \,, 
\end{align}
which is the product of two pieces that are separately $\mu$-independent. We
refer to this procedure as ``split matching'' because formally we match diagrams
at two scales  rather than at a single scale. The two matching $\mu$'s are
``split'' because they are parametrically similar in the power-counting sense.

We organize the remainder of our paper as follows. We begin by using split
matching to determine the hard matching functions, ${\cal B}=B_1 B_2$, for $B\to
X_s \ell^+\ell^-$ in SCET; this is one of the main points of our paper.  It is
discussed in Sec.~\ref{match} at leading power and one-loop order (including
both bottom-, charm-, and light-quark loops and other virtual corrections). The
extension to higher orders is also illustrated. Steps i) and ii) are summarized
in Sec.~\ref{match}, together with Appendix~\ref{app:defs}.  In Sec.~\ref{Run1}
we discuss the running for step iv) and give a brief derivation of why the
anomalous dimension is independent of the Dirac structure to all orders in
$\alpha_s$.  In Sec.~\ref{decay}, we discuss the basic ingredients for the
triply differential decay rate and the forward-backward asymmetry in terms of
hadronic tensors. A second main point of our paper is the SCET matrix-element
computation for $B\to X_s\ell^+\ell^-$, step v), which is performed in
Sec.~\ref{SCET}.  In Sec.~\ref{Run} we review the running for the shape
function, step vi).  In Sec.~\ref{results} we present our final results for the
differential decay rates at leading order in the power expansion, including all
the ingredients from Sec.~\ref{LOanalysis} and incorporating the relevant
experimental cuts. The triply differential spectrum and doubly differential
spectra are derived in subsections~\ref{results}A-D.  Readers interested only in
our final results may skip directly to section~\ref{results}.  We compare
numerical results for matching coefficients at $m_b$ with terms in the local OPE
in Sec.~\ref{numbers}.  In Appendix~\ref{App:q2jet} we briefly comment on how
our analysis will change if we assume a parametrically small dilepton invariant
mass, $q^2\sim \lambda^2$, rather than the scaling $q^2\sim \lambda^0$ used in
the body of the paper.  (For the case $q^2\sim \lambda^2$, the rate for $B\to
X_s\ell^+\ell^-$ would {\em not} be determined by a factorization theorem with
the same structure as for $B\to X_u\ell\bar\nu$.)

 
\section{Analysis in the Shape Function Region
} \label{LOanalysis}

\subsection{Matching on to SCET} \label{match}

We begin by reviewing the form of the electroweak Hamiltonian obtained after
evolution down to the scale $\mu\simeq m_b$, and then perform the leading-order
matching of this Hamiltonian on to operators in SCET. For the treatment of
$\gamma_5$ we use the NDR scheme throughout.  Below the scale $\mu=m_W$, the
effective Hamiltonian for $b\to s\ell^+\ell^-$ takes the form~\cite{gsw89}
\begin{eqnarray}
{\cal H}_{W} & = & -\, \frac{4 G_F}{\sqrt{2}} V_{tb} V_{ts}^*
        \sum_{i=1}^{10} C_i (\mu)  {\cal O}_i(\mu) \, ,
   \label{eq:H_eff}
\end{eqnarray}
where we have used unitarity of the CKM matrix to remove $V_{cb}V^*_{cs}$
dependence and have neglected the tiny $V_{ub}V^*_{us}$ terms.  The operators
${\cal O}_i(\mu)$ are
\begin{align}
  {\cal O}_1 &= (\bar{s}_{L \alpha} \gamma_\mu b_{L \beta})
  (\bar{c}_{L \beta} \gamma^\mu c_{L \alpha}),   
  & {\cal O}_2 &= (\bar{s}_{L \alpha} \gamma_\mu b_{L \alpha})
  (\bar{c}_{L \beta} \gamma^\mu c_{L \beta}),   \\
  {\cal O}_3 &= (\bar{s}_{L \alpha} \gamma_\mu b_{L \alpha}) \sum_{q=u,d,s,c,b}
  (\bar{q}_{L \beta} \gamma^\mu q_{L \beta}),   
  & {\cal O}_4 &= (\bar{s}_{L \alpha} \gamma_\mu b_{L \beta}) \sum_{q=u,d,s,c,b}
  (\bar{q}_{L \beta} \gamma^\mu q_{L \alpha}),   \nn \\
  {\cal O}_5 &= (\bar{s}_{L \alpha} \gamma_\mu b_{L \alpha}) \sum_{q=u,d,s,c,b}
  (\bar{q}_{R \beta} \gamma^\mu q_{R \beta}),   
  &{\cal O}_6 &= (\bar{s}_{L \alpha} \gamma_\mu b_{L \beta}) \sum_{q=u,d,s,c,b}
  (\bar{q}_{R \beta} \gamma^\mu q_{R \alpha}),   \nn \\
  {\cal O}_7 &= \frac{e}{16 \pi^2} \bar{s} \sigma_{\mu \nu}F^{\mu \nu}
   (\bar m_b P_R + \bar m_s P_L) b , 
  &{\cal O}_8 &= \frac{g}{16 \pi^2} \bar{s}_{\alpha} T_{\alpha \beta}^a
  \sigma_{\mu \nu} (\bar  m_b P_R + \bar m_s P_L)
  b_{\beta} G^{a \mu \nu},  \nn \\
  {\cal O}_9 &= \frac{e^2}{16 \pi^2} \bar{s}_{L\alpha} \gamma^{\mu} b_{L \alpha}
  \bar{\ell} \gamma_{\mu} \ell , 
  &{\cal O}_{10} &= \frac{e^2}{16 \pi^2} \bar{s}_{L \alpha} \gamma^{\mu} b_{L
    \alpha} \bar{\ell} \gamma_{\mu}\gamma_5 \ell \,, \nn
\end{align}
where $P_{R,L}=(1\pm \gamma_5)/2$.  In the following, we shall neglect the mass
of the strange quark in ${\cal O}_{7,8}$. For our analysis, $m_s$ is not needed
as a regulator for IR divergences, which are explicitly cut off by
non-perturbative scales $\sim \Lambda_{\rm QCD}$. In the shape function region,
the $m_s$ dependence is small and was computed in Ref.~\cite{Leibms}.
Non-perturbative sensitivity to $m_s$ shows up only at subleading power, while
computable ${\cal O}(m_s^2/m_b\Lambda_{\rm QCD})$ jet-function corrections are
numerically smaller than the $\Lambda_{\rm QCD}/m_b$ power corrections.

At NLL order, one requires the NLL Wilson coefficient of ${\cal O}_9$ and the LL
coefficients of the other operators. For ${\cal O}_{7,9,10}$ these are given
by~\cite{misiak93,bm95}
\begin{eqnarray} \label{eq:C7910}
C_7^{\rm NDR}(\mu) & = & r_0^{-\frac{16}{23}}\: C_7(M_W) 
   + \frac{8}{3} \left(r_0^{-\frac{14}{23}} - r_0^{-\frac{16}{23}}\right) 
   C_8(M_W) + \sum_{i=1}^8 t_i\: r_0^{-a_i}, \\
C_9^{\rm NDR}(\mu) & = &
    P_0^{\rm NDR}(\mu) + \frac{Y(m_t^2/M_W^2)}{\sin^2\theta_W} 
    -4 Z(m_t^2/M_W^2) + P_E(\mu) E(m_t^2/M_W^2), \nn \\
C_{10}(\mu)  & = & C_{10}(M_W)= - \frac{Y(m_t^2/M_W^2)}{\sin^2\theta_W} \,, \nn 
\end{eqnarray}
where $C_7(m_W)$, $C_8(m_W)$ and the Inami-Lim functions $Y$, $Z$, and $E$ are
obtained from matching at $\mu=m_W$, and are given in Appendix~\ref{app:defs}.
The $\mu$-dependent factors include~\cite{misiak93,bm95}
\begin{eqnarray}
P_0^{\rm NDR}(\mu) & = & \frac{\pi}{\alpha_s(M_W)} \left(-0.1875 + \sum_{i=1}^8 p_i\:
r_0^{-a_i-1}\right) + 1.2468 +  \sum_{i=1}^8 r_0^{-a_i} \left(
\rho^{\rm NDR}_i + s_i\, r_0^{-1} \right) \,, \nn \\
P_E(\mu) & = & 0.1405 +\sum_{i=1}^8 q_i\: r_0^{-a_i-1} \,, \qquad\quad
r_0  =  \frac{\alpha_s(\mu)}{\alpha_s(m_W)} \,.
\end{eqnarray}
The numbers $t_i$, $a_i$, $\rho_i^{\rm NDR}$, $s_i$, $q_i$ that appear here are
listed in Appendix~\ref{app:defs}. Results for the running coefficients of
the four-quark operators, $C_{1-6}(\mu)$, can be found in Ref.~\cite{bm95}. We
have modified the standard notation slightly (e.g. $r_0(\mu)$) to conform with
additional stages of the RG evolution discussed in sections~\ref{Run1} and
\ref{Run}. Contributions beyond NLL will be mentioned below.

At a scale $\mu \approx m_b$, we need to match $b\to s\ell^+\ell^-$ matrix
elements of ${\cal H}_{W}$ on to matrix elements of operators in SCET with a
power expansion in the small parameter $\lambda$, where $\lambda^2 =
\Lambda_{\rm QCD}/m_b$. For convenience, we refer to the resulting four-fermion
scalar operators in SCET as ``currents'' and use the notation $J_{\ell\ell}$. In
SCET we also need the effective Lagrangians. The heavy quark in the initial
state is matched on to an HQET field $h_v$, and the light energetic strange
quark is matched on to a collinear field $\xi_n$. For the leading-order analysis
in $\Lambda/m_b$  we need only the lowest-order terms,
\begin{eqnarray}
  {\cal H}_W & = &
  -\,\frac{G_F \alpha}{\sqrt{2} \pi} \,( V_{tb} V_{ts}^*)  
  \: {J}^{(0)}_{\ell\ell}\,,
\qquad
  {\cal L} = {\cal L}^{(0)}_{\rm HQET} + {\cal L}^{(0)}_{\rm SCET} \,, 
\end{eqnarray}
where ${J}^{(0)}_{\ell\ell}$ is the LO operator and the quark contributions
to the HQET and SCET actions are
\begin{eqnarray} \label{LLO}
  {\cal L}_{\rm HQET}^{(0)} &=& \bar h_v iv\mcdot D_{us} h_v \,, \nn\\
  {\cal L}_{\rm SCET}^{(0)} &=& \bar \xi_n \left[ i n\mcdot {D}_c
  + i \Dslash^\perp_c   \frac{1}{i\bn\mcdot D_c}
  i \Dslash^\perp_c\right] \frac{\bnslash}{2} \: \xi_n  \,. 
\end{eqnarray}
The covariant derivatives $D_{us}$ and $D_c$ involve ultrasoft and collinear
gluons respectively, and we have made a field redefinition on the collinear
fields to decouple the ultrasoft gluons at LO~\cite{bps02}. For convenience, we
define the objects
\begin{eqnarray} \label{calB}
  {\cal H}_v &=& Y^\dagger h_v\,,\qquad
   \psi_{us} = Y^\dagger q_{us} \,, \qquad\qquad
    {\cal D}_{us} = Y^\dagger D_{us} Y 
  \nn\\
    \chi_n &=& W^\dagger \xi_n \,, \qquad
  {\cal D}_c = W^\dagger D_c W \,,\qquad
  ig {\cal B}_c^\mu
   = \Big[\frac{1}{\bnP} W^\dagger  [i\bn\mcdot D_c , iD_c^\mu] W\Big] \,,
\qquad 
\end{eqnarray}
which contain ultrasoft and collinear Wilson lines, 
\begin{eqnarray}
Y(x) & = & P\exp\Big(ig \int_{-\infty}^0\!\!\!\! ds\:
     n\mcdot A_{us}(x\!+\! ns)\Big) 
\end{eqnarray}
and
\begin{eqnarray} \label{W}
W(x) & = & P\exp\Big(ig \int_{-\infty}^0\!\!\!\! ds\:
     \bn\mcdot A_{n}(x\!+\! s\bn)\Big) \,,
\end{eqnarray}
as well as the label operator $\bnP$~\cite{bs1}.

To simplify the analysis we treat both $m_c$ and $m_b$ as hard scales and 
integrate out both charm and bottom loops at $\mu\simeq m_b$.  At leading 
order in SCET, the currents that we match on to are
\begin{eqnarray} \label{JscetLOold}
 J_{\ell\ell}^{(0)} &=& \sum_{i=a,b,c} C_{9i}(s)\: 
     \big(\bar\chi_{n,p} \Gamma_{i}^{(v)\mu} {\cal H}_v\big)
     \big( \bar{\ell} \gamma_{\mu} \ell \big)
  + \sum_{i=a,b,c} C_{10i}(s)\: 
     \big( \bar\chi_{n,p} \Gamma_{i}^{(v)\mu} {\cal H}_v\big)
     \big( \bar{\ell} \gamma_{\mu} \gamma_5 \ell \big) \nn\\
&& - \sum_{j=a,\ldots,d} C_{7j}(s)\: 2 m_B
      \big( \bar\chi_{n,p} \Gamma_{j}^{(t)\mu} {\cal H}_v \big)
      \big( \bar{\ell} \gamma_{\mu} \ell \big) \,, 
\end{eqnarray}
where the sum is over Dirac structures to be discussed below. The simple
structure of these LO SCET operators is quite important to our analysis: for
example, by power counting there are no four-quark operators that need to be
included in SCET at this order.  In Eq.~(\ref{JscetLOold}) two auxiliary
four-vectors appear, $v^\mu$ and $n^\mu$.  The $B$ momentum, total momentum of
the leptons, and jet momentum (sum of the four-momenta of all the hadrons in
$X_s$) are
\begin{align}
& p_B^\mu = m_B v^\mu \,, 
 & q^\mu & = p_{\ell^+}^\mu + p_{\ell^-}^\mu \,,
 & p_X^\mu & = n\mcdot p_X \: \frac{\bn^\mu}{2} 
    + \bn\mcdot p_X \: \frac{n^\mu}{2}  \,,
\end{align}
respectively. Here $v^2=1$ and $n^\mu$ and $\bn^\mu$ are light-like vectors,
which satisfy $n^2 = \bn^2 = 0$ and $n\cdot\bn=2$.  The components of a vector
can then be written as $(p^+,p^-,p_\perp) = (n\cdot p, \bn\cdot p,
p_\perp^\mu)$.  We use a frame in which $q_\perp^\mu = v_\perp^\mu = 0$ and
$v^\mu = (n^\mu + \bn^\mu)/2$. Since $p_X = m_B v-q$ we have
\begin{align}
 & p_X^2 = m_X^2= \bn\mcdot p_X \: n\mcdot p_X  = m_B^2 +q^2 
    - m_B(n\mcdot q\!+\! \bn\mcdot q) \,,\qquad
  q^2 = \bn\mcdot q \: n\mcdot q \,,\nn\\
 & \bn\mcdot p_X = m_B -\bn\mcdot q \,,\qquad 
   n\mcdot p_X = m_B - n\mcdot q \,.
\end{align}  
For later convenience we define the hadronic dimensionless variables
\begin{eqnarray} \label{ybaru}
  x_H = \frac{2 E_{\ell^-}}{m_B} \,, \qquad
  \overline y_H = \frac{\bn\mcdot p_X}{m_B} \,,\qquad 
  u_H = \frac{n\mcdot p_X}{m_B} \,, \qquad
  y_H = \frac{q^2}{m_B^2} \,.
\end{eqnarray}

In SCET the total partonic $\bn\cdot p$ momentum of the jet is a hard momentum
$\sim m_b$ and also appears in the SCET Wilson coefficients.  At LO,
$\bn\mcdot p = (m_b^2 - q^2)/m_b$ and demanding that $\bn\mcdot p$ is large
means only that $q^2$ cannot be too close to $m_b^2$. For example, neither
$q^2\approx 0$ nor $q^2\approx m_b^2/2$ modifies the power counting for
$\bn\mcdot p$. Thus, there is no requirement to impose a scaling that $q^2$ be
small. For convenience, in the hard coefficients we write
\begin{align}
 &  C\big( {\bn\mcdot p},{m_b}, {\mu_0}, {\mu_b} \big)
   \rightarrow C\big(s, m_b, \mu_0, \mu_b \big) \,, 
  & s = \frac{q^2}{m_b^2} \,,
\end{align}
since the partonic variable $s$ is a more natural choice in $b\to s\ell^+\ell^-$
and is equivalent at LO. For purposes of power counting in this paper we count
$s\sim \lambda^0$.  We shall see in section~\ref{numbers} that varying $s$
causes a very mild change in the coefficients. In Appendix~\ref{App:q2jet} we
briefly explore a different scenario, in which $s\sim \lambda^2$.  A
distinction between two matching scales $\mu_0$ and $\mu_b$ is made in $C$ in
order to separate the decay rate into two $\mu$-independent pieces, as displayed
in Eq.~(\ref{mu2}).  For power counting purposes, $\mu_0\sim \mu_b\sim m_b$ and
formally $\mu_0 \ge \mu_b$.  For numerical work one can take $\mu_0=\mu_b$.

In Eq.~(\ref{JscetLOold}) we begin with a complete set of Dirac structures for the
vector and tensor currents in SCET, namely
\begin{eqnarray} \label{Jeff3}
 \Gamma_{a-c}^{(v)} &=&  P_R \Big\{ \gamma^\mu\,, v^\mu \,,
   \frac{n^\mu}{n\mcdot v} \Big\}\,,\quad
 \Gamma_{a-d}^{(t)} =  P_R\  \frac{q_\tau}{q^2}\: \Big\{ i\sigma^{\mu\tau} \,,
   \gamma^{[\mu} v^{\tau]}\,, \frac{\gamma^{[\mu} n^{\tau]} }{n\mcdot v} \,,
   \frac{n^{[\mu}v^{\tau]}}{n\mcdot v}  \Big\} \,. 
\end{eqnarray}
These come with Wilson coefficients $C_{9a,b,c}$ and $C_{7a,b,c,d}$ respectively.
This basis is over-complete for $B\to X_s\ell^+\ell^-$, but considering a
redundant basis makes it easy to incorporate pre-existing perturbative
calculations for the currents into our computations. Only the coefficients
$C_{7a,9a}$ appear at tree level, but for heavy-to-light currents it is known
that the other structures become relevant once perturbative corrections are
included.  For simplicity of notation, we treat the $1/q^2$ photon propagator in
$\Gamma^{(t)}_{j}$ as part of the effective-theory operator.\footnote{ If we
  instead demand that the momentum $q^2$ be collinear in the $\bn$ direction,
  with $s\sim \lambda^2$, then the SCET operator with a photon field strength
  should be kept, and will then be contracted with an operator with collinear
  leptons within SCET. In this case there will also be additional four-quark
  operators needed in the basis in Eq.~(\ref{JscetLO}). }

To reduce the basis in Eq.~(\ref{Jeff3}) further, we can use i) current
conservation, $q^\mu \bar \ell \gamma_\mu \ell=0$, ii) $q^\mu \bar \ell
\gamma_\mu \gamma_5 \ell=0$ for massless leptons, iii) a reduction of the tensor
$\Gamma^{(t)}$ Dirac structures into vector structures, since they are all
contracted with $q_\tau$.  Constraint ii) allows us to eliminate $C_{10c}$.
Taken together, constraints i) and iii) allow us to reduce the seven terms
$C_{9i}$ and $C_{7i}$ to two independent coefficients.  For our new basis
of operators we take
\begin{eqnarray} \label{JscetLO}
 J_{\ell\ell}^{(0)} &=&  {\cal C}_{9}\: 
     \big(\bar\chi_{n,p} P_R \gamma^\mu {\cal H}_v\big)
     \big( \bar{\ell} \gamma_{\mu} \ell \big) 
  -  {\cal C}_7\:   \frac{2 m_B\, q_\tau}{q^2}
      \big( \bar\chi_{n,p} P_R \, i\sigma^{\mu\tau} {\cal H}_v \big)
      \big( \bar{\ell} \gamma_{\mu} \ell \big)\nn\\[4pt]
  && {} + {\cal C}_{10a}\: 
     \big( \bar\chi_{n,p} P_R \gamma^\mu {\cal H}_v\big)
     \big( \bar{\ell} \gamma_{\mu} \gamma_5 \ell \big) 
    + {\cal C}_{10b}\: 
     \big( \bar\chi_{n,p} P_R v^\mu {\cal H}_v\big)
     \big( \bar{\ell} \gamma_{\mu} \gamma_5 \ell \big)\,,
\end{eqnarray}
and find that
\begin{align}\label{switchbasis}
  {\cal C}_{9} &=  C_{9a}  + \frac{C_{9b}}{2} 
   - \frac{m_B }{ n\mcdot q}\, C_{7b}
   + \frac{2m_B(C_{7c}-C_{7d})+n\mcdot q\, C_{9c}}{n\mcdot q \!-\!\bn\mcdot q}  
 \,,\\
  {\cal C}_{7} &=  C_{7a} - \frac{C_{7b}}{2} 
     - \frac{\bn\mcdot q}{4 m_B} C_{9b} +  
   \frac{1}{n\mcdot q \!-\!\bn\mcdot q} \Big[ \frac{-q^2}{2m_B}\, C_{9c} 
      - n\mcdot q\, C_{7c} + \bn\mcdot q \, C_{7d} \Big] \,,\nn\\
  {\cal C}_{10a} &=  C_{10a} \,, \nn \\
  {\cal C}_{10b} &=  C_{10b} 
     + \frac{2\,n\mcdot q }{n\mcdot q \!-\!\bn\mcdot q} \, C_{10c} \,.\nn
\end{align}
Our Dirac structures for the ${\cal C}_9$ and ${\cal C}_7$ terms in
Eq.~(\ref{JscetLO}) were deliberately chosen, in order to make results for the
decay rates appear as much as possible like those in the local OPE. The fact
that the basis of SCET operators for $B\to X_s\ell^+\ell^-$ involves only
bilinear hadronic currents at LO means that in the leading-order factorization
theorem we find the exact same non-perturbative shape function as for $B\to
X_s\gamma$ and $B\to X_u\ell\bar\nu$. This is immediately evident from the
operator-based proof of factorization in Ref.~\cite{bps02}, for example. While
the coefficients $C_{9i}$, $C_{7i}$, $C_{10i}$ in Eq.~(\ref{JscetLOold}) are
functions only of $s = (n\mcdot q) \, (\bn\mcdot q) / m_b^2$, the reduction of
the basis of operators brings in additional kinematic dependence on $\bn\mcdot
q$ and $n\mcdot q$ for the ${\cal C}_i$'s (which is also the case in analyzing
exclusive dilepton decays~\cite{Grinstein:2005ud}).  At tree level we have
${\cal O}_{9,10}$ contributing to $C_{9a}$ and $C_{10a}$, and a contribution
from ${\cal O}_7$ with the photon producing an $\ell^+\ell^-$ pair, which give
\begin{align} \label{treematch}
  & {\cal C}_{9}= C_9^{\rm NDR}(\mu_0)\,, 
  & {\cal C}_{7} & =
    \frac{\overline m_b(\mu_0)}{m_B}\, C_7^{\rm NDR}(\mu_0) \,,
  & {\cal C}_{10a} = C_{10} \,,\qquad
  & {\cal C}_{10b} = 0 \,.
\end{align}
Beyond tree level there will be $C_7$ dependence in ${\cal C}_9$, and $C_9$
dependence in ${\cal C}_7$. Eq.~(\ref{treematch}) indicates that with our choice
of basis the same short-distance dependence dominates in SCET: ${\cal
  C}_9\approx C_9$, etc. We explore this further in Section~\ref{numbers}. In
Eq.~(\ref{treematch}) there is no distinction as to whether this matching is
done at $\mu=\mu_0$ or $\mu=\mu_b$. The effective-theory operator in
Eq.~(\ref{JscetLO}) was defined with a factor of $m_B$ pulled out so that the
$\mu$-dependent factors $\overline m_b C_7^{\rm NDR}$ are contained in the
coefficients ${\cal C}_{7}$.

\begin{figure}[t!]
 \vskip .2cm
 \centerline{
  \mbox{\epsfysize=3.5truecm \hbox{\epsfbox{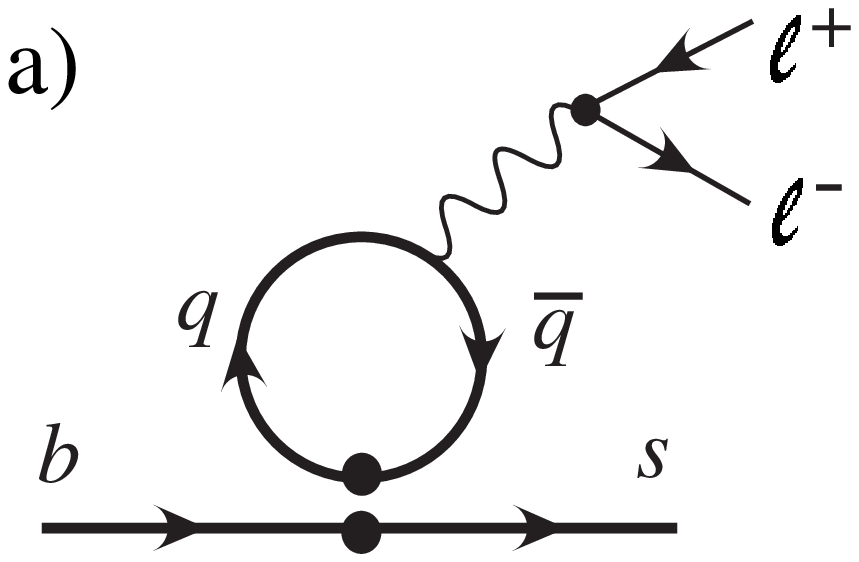}} } \qquad\qquad
  \mbox{\epsfysize=3.5truecm \hbox{\epsfbox{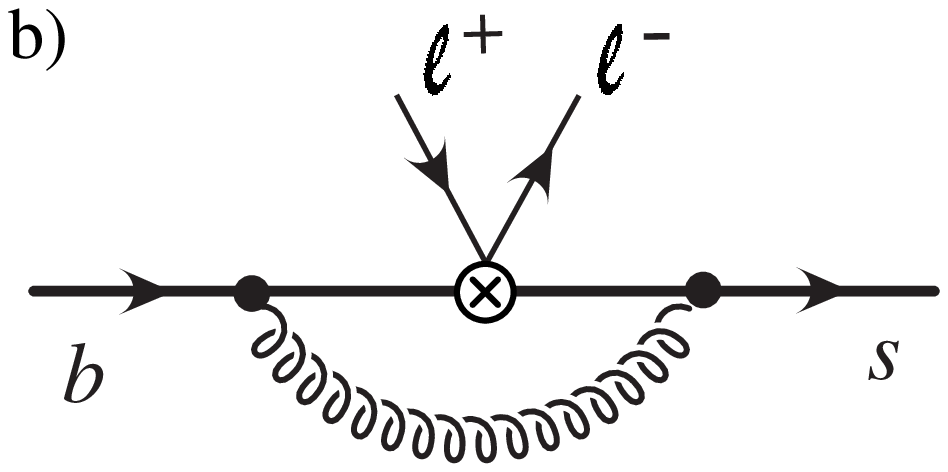}} }
  }
 \vskip .2cm
 {\caption[1]{Graphs from $H_W$ for matching on to SCET.}
\label{fig_match} }
 \vskip -0.1cm
\end{figure}

\begin{figure}[t!]
 \vskip .2cm
 \centerline{
  \mbox{\epsfysize=4.truecm \hbox{\epsfbox{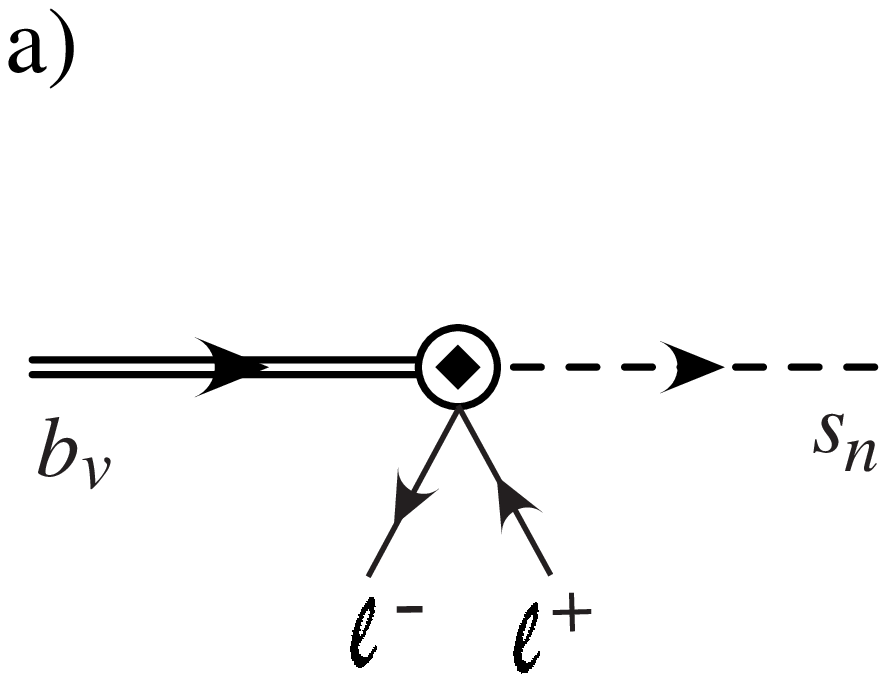}} } \quad
  \mbox{\epsfysize=4.truecm \hbox{\epsfbox{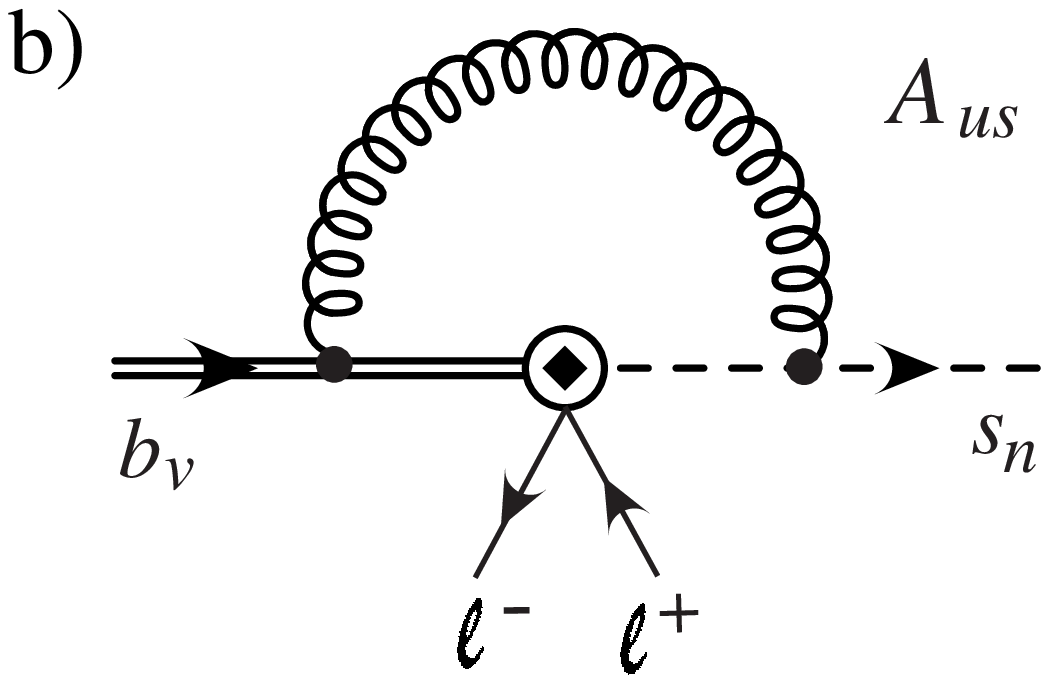}} } \quad
  \mbox{\epsfysize=4.truecm \hbox{\epsfbox{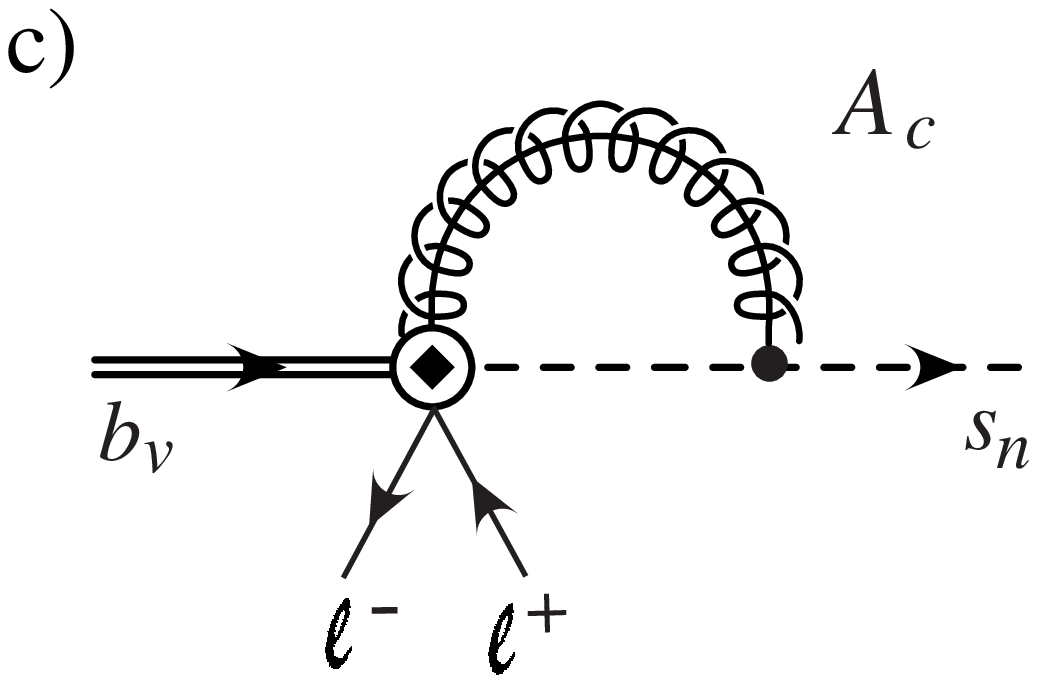}} }
  }
 \vskip -.4cm
 {\caption[1]{Graphs in SCET for the matching computation.}
\label{fig_scetm} }
 \vskip -0.1cm
\end{figure}
At one-loop order, the full-theory diagrams needed for the matching are shown in
Fig.~\ref{fig_match} (plus wave-function renormalization, which is not shown).
At this order the four-quark operators ${\cal O}_{1-6}$ contribute through
Fig.~\ref{fig_match}a.  The one-loop graphs in SCET with the operators in
Eq.~(\ref{JscetLO}) are shown in Fig.~\ref{fig_scetm} (plus wave-function
renormalization, which is not shown).  There are no graphs with four-quark
operators within SCET since we treat $q^2\sim \lambda^0$, so
Fig.~\ref{fig_match}a matches directly on to ${\cal C}_{9}$.

As discussed in the introduction, we perform a split-matching procedure from the
full theory above $m_b$ on to SCET below $m_b$, making use of two matching scales
$\mu_0$ and $\mu_b$. Contributions from this stage of matching therefore take
the form
\begin{align} \label{B1B2}
 {\cal B}(\mu_0,\mu_b) = B_1(\mu_0) B_2(\mu_b) \,.
\end{align}  
Since ${\cal O}_{10}$ has no anomalous dimension above $m_b$ and there is a
common universal anomalous dimension for all the operators in
$J^{(0)}_{\ell\ell}$ below $m_b$, there is a well-defined prescription for carrying
this out. We take all contributions that cause perturbative corrections to
${\cal C}_{10a}$ and ${\cal C}_{10b}$ to be at the scale $\mu_b$, so for this
operator $B_1(\mu_0)=C_{10}$, and at one-loop order $B_2(\mu_b)$ includes
$\alpha_s(\mu_b)\ln^2(\mu_b)$, $\alpha_s(\mu_b)\ln(\mu_b)$, and
$\alpha_s(\mu_b)$ terms from matching the vertex diagram Fig.~\ref{fig_match}b
and wave-function diagrams on to SCET. The analogous contributions from vertex
diagrams for ${\cal C}_9$ and ${\cal C}_7$ are also matched at $\mu=\mu_b$ to
determine their $B_2(\mu_b)$'s (for ${\cal C}_7$ the full-theory tensor current
has a $\ln\mu$ that is matched at $\mu=\mu_0$).  The universality of the
anomalous dimensions in SCET guarantees that this procedure remains well defined
at any order in perturbation theory and can be organized into the product
structure displayed in Eq.~(\ref{B1B2}). For ${\cal C}_9$ and ${\cal C}_7$ there
are additional non-vertex-like contributions that are matched on to $B_1(\mu_0)$
at a scale $\mu_0 \ge \mu_b$.  These include contributions from four-quark
operators ${\cal O}_{1-6}$ in the full theory, which will match on to ${\cal
  C}_9$ and ${\cal C}_7$ in SCET.

The difference between the full-theory diagram in Fig.~\ref{fig_match}b and the
SCET graphs in Fig.~\ref{fig_scetm}b,c is IR finite (where we must use the same
IR regulator in both theories, as is always the case for matching computations).
In the UV the full-theory graph in Fig.~\ref{fig_match}b plus wave-function
renormalization is $\mu$-independent since the current is conserved. The graphs
in SCET induce a $\mu$ dependence and an anomalous dimension for the
effective-theory currents. These terms are matched at $\mu=\mu_b$. We start with
the basis in Eq.~(\ref{JscetLOold}) and find
\begin{eqnarray} \label{match1}
C_{10a}(\mu_0,\mu_b) & = & C_{10} \Big[ 1 +
   \frac{\alpha_s(\mu_b)}{\pi}\: \omega_a^V(s,\mu_b) \Big]\,, \nn \\
C _{10b,10c}(\mu_0,\mu_b) & = & C_{10} \, \frac{\alpha_s(\mu_b)}{\pi}\: 
 \omega_{b,c}^V(s) \,,
\end{eqnarray}
with a constant $\mu_0$-independent $C_{10}$.  The perturbative coefficients
were computed in Ref.~\cite{bfps01}, and setting $\bn\mcdot p/m_b = (1-s)$ we find
\begin{align}
\omega_a^V(s,\mu_b) & =  - \frac{1}{3} \Big[ 2\!\ln^2(1\!-\!s)
  + 2 {\rm Li}_2(s)
  + \ln(1\!-\! s) \Big( \frac{1 \!-\! 3 s}{s}\Big)
  + \frac{\pi^2}{12} + 6 \nn\\
  &\qquad\quad  +2 \ln^2\Big(\frac{\mu_b}{m_b}\Big) 
    +5 \ln\Big(\frac{\mu_b}{m_b}\Big)  
    - 4\ln(1\!-\!s)\ln\Big(\frac{\mu_b}{m_b}\Big) \
   \Big] \,, \nn \\[4pt]
\omega_b^V(s) & =  \frac{1}{3} \Big[  \frac{2}{s}
  + \frac{2(1-s)} {s^2} \ln(1-s)\Big] \,, \nn \\[4pt]
\omega_c^V(s) & = \frac{1}{3} \Big[ \frac{(2 s - 1)(1-s)}{s^2}\ln(1-s)
  - \frac{(1-s)}{s}  \Big] \,.
\end{align}

For the matching on to $C_{9a,b,c}$ in the basis in Eq.~(\ref{JscetLOold}) we have
the same perturbative coefficients $\omega_{a,b,c}$ as for $C_{10a,b,c}$, because
only the leptonic current differs:
\begin{eqnarray} \label{C9match}
C_{9a}(\mu_0,\mu_b) & = & C_9^{\rm mix}(\mu_0) \Big[ 1 +
   \frac{\alpha_s(\mu_b)}{\pi}\: \omega_a^V(s,\mu_b) \Big] 
\,,\nn \\
C _{9b,9c}(\mu_0,\mu_b) & = & C_9^{\rm mix}(\mu_0)
  \,\Big[ \frac{\alpha_s(\mu_b)}{\pi}\: 
  \omega_{b,c}^V(s) \Big]
\,.
\end{eqnarray}
However, for $C_{9i}$ there are additional contributions, $C_9^{\rm mix}(\mu_0)$,
from the matching at $\mu=\mu_0$, which at one-loop order and ${\cal O}(\alpha_s^0)$
includes Fig.~\ref{fig_match}a:
\begin{align} \label{C9mix}
 C_9^{\rm mix}(\mu_0) &=  C_9^{\rm NDR}(\mu_0)
 + \frac{2}{9} \left( 3 C_3 + C_4 + 3 C_5 + C_6 \right)
 - \frac{1}{2} h(1, s) \left( 4 C_3 + 4 C_4 + 3 C_5 + C_6 \right)
 \nn \\
&  + h\big(\frac{m_c}{m_b}, s\big)\left( 3 C_1 + C_2 + 3 C_3 + C_4 + 3
C_5 + C_6 \right) - \frac{1}{2} h(0, s) \left( C_3 + 3 C_4 \right)  \nn \\
 & + \frac{\alpha_s(\mu_0)}{\pi}\: C_9^{\rm mix (1)}(\mu_0) \,,
\end{align}
where all running coefficients on the RHS are $C_i=C_i(\mu_0)$. We shall discuss
the relation of $C_9^{\rm mix}$ to $\tilde C_9^{\rm eff}$ in in the local OPE
analysis~\cite{misiak93,bm95} after Eq.~(\ref{Cresult}). In Eq.~(\ref{C9mix}) the
functions $h(1,s)$, $h(z,s)$, and $h(0,s)$ for the b-quark, c-quark, and
light-quark penguin loops are~\cite{gsw89,misiak93}
\begin{eqnarray}
h(z,s) &=& \frac{8}{9}\ln (\frac{\mu_0}{m_b})
  -\frac{8}{9} \ln z + \frac{8}{27} +\frac{4}{9}\zeta
 -\frac{2}{9}(2 + \zeta) \sqrt{|1-\zeta|} \nn\\
& & \times \bigg[
 \theta(1-\zeta)\bigg(-i\pi+\ln\frac{1+\sqrt{1-\zeta}}{1-\sqrt{1-\zeta}} 
   \bigg) +\theta(\zeta-1)\: 2 \arctan \frac{1}{\sqrt{\zeta-1}} \bigg] 
 \,, \nn \\
h(0,s) & = & \frac{8}{27}+\frac{8}{9}\ln (\frac{\mu_0}{m_b})
              -\frac{4}{9}\ln s + \frac{4}{9}i\pi \,, 
\end{eqnarray}
with $\zeta = 4 z^2/s$.  Higher-order ${\cal O}(\alpha_s)$ corrections in
Eq.~(\ref{C9mix}) are denoted by the $C_9^{\rm mix (1)}$ term. An important
class of these corrections from mixing can be determined from the NNLL analysis
in Refs.~\cite{aagw02,ghiy03,ghiy04}:
\begin{align}
 C_9^{\rm mix (1)}(\mu_0) &=   C_8^{\rm NDR}\,
 \kappa_{8\to 9}(s,\mu_0) 
 + C_1\, \kappa_{1\to 9}(s,\mu_0,\hat m_c)
   + C_2 \, \kappa_{2\to 9}(s,\mu_0,\hat m_c) 
\,. 
\end{align}
To determine these terms one must be careful to separate out the factors in
square brackets in Eq.~(\ref{C9match}). However we shall not attempt to include
all NNLL terms consistently here. Contributions to $C_{9}^{\rm mix (1)}$ from
the penguin coefficients $C_{3-6}$ are unknown but expected to be small (at the
$\sim 1\%$ level).

Lastly, we turn to the results for $C_{7i}$. From the vertex graphs we have
\begin{align}
C_{7a}(\mu_0,\mu_b) & =  
  C_7^{\rm mix}(\mu_0)\: \Big[  1 + \frac{\alpha_s(\mu_b)}{\pi}\: \omega^T_a(s,\mu_b)
    \Big] 
    \,,\nn \\
C_{7b,7c,7d}(\mu_0,\mu_b) & =   
  C_7^{\rm mix}(\mu_0)\: \frac{\alpha_s(\mu_b)}{\pi}\: \omega^T_{b,c,d}(s) \,.
\end{align}
The $\omega_i^T$ perturbative corrections are again determined from
the SCET matching in Ref.~\cite{bfps01}, which (switching to $s$) gives
\begin{align} \label{kappa}
\omega^T_a(s,\mu_b) & =  - \frac{1}{3} \Big[ 2\!\ln^2(1\!-\!s)
  + 2 {\rm Li}_2(s)
  + \ln(1\!-\! s) \Big( \frac{2 \!-\! 4 s}{s}\Big)
  + \frac{\pi^2}{12} + 6 \nn\\
  &\qquad\quad  +2 \ln^2\Big(\frac{\mu_b}{m_b}\Big) 
    +5 \ln\Big(\frac{\mu_b}{m_b}\Big)  
    - 4\ln(1\!-\!s)\ln\Big(\frac{\mu_b}{m_b}\Big) \
   \Big] \,, \nn \\[4pt]
\omega^T_b(s) & = \omega^T_d(s) = 0 \,, \nn\\[4pt]
\omega^T_c(s) &= \frac{1}{3} \Big[ \frac{-2(1-s)\ln(1-s)}{s}
    \Big] \,.
\end{align}
Additional contributions from other diagrams are matched at the scale $\mu_0$
into $C_7^{\rm mix}(\mu_0)$.  Note that, unlike the vector currents, the tensor
current for $O_7$ gets renormalized for $\mu > m_b$ and we must include the
corresponding $\ln(\mu_0/m_b)$ in $C_7^{\rm mix}(\mu_0)$, i.e.
\begin{align} \label{C7mix}
  C_7^{\rm mix}(\mu_0) &=  \frac{\overline m_b(\mu_0)}{m_B} \: \bigg\{
  C_7^{\rm NDR}(\mu_0)\: \Big[ 1 - \frac{2\alpha_s(\mu_0)}{3\pi}\:
  \ln\Big(\frac{\mu_0}{m_b}\Big)  \Big]\nn\\
 &\qquad\qquad\qquad\quad
  + \frac{\alpha_s(\mu_0)}{\pi}\: C_7^{\rm mix(1)}(\mu_0) \bigg\} \,,
\end{align}
where, much like in the case of $C_9^{\rm mix}$, we have
\begin{align}
  C_7^{\rm mix(1)}(\mu_0) &=  C_8^{\rm NDR}\, \kappa_a^8(s,\mu_0) +
  C_1 \, \kappa_a^1(s,\mu_0,\hat m_c) + C_2 \, \kappa_a^2(s,\mu_0,\hat m_c) \,,
\end{align}
and the results for $\kappa_{8\to 7}(s,\mu_0)$, $\kappa_{1\to 7}(s,\mu_0,\hat
m_c)$, and $\kappa_{2\to 7}(s,\mu_0,\hat m_c)$ can be found in
Ref.~\cite{greub96}.  Contributions to $C_{7}^{\rm mix (1)}$ from the penguin
coefficients $C_{3-6}$ can be found in Ref.~\cite{Buras:2002tp}.

Using Eq.~(\ref{switchbasis}), $\bn\mcdot q\, n\mcdot q/m_B^2 =y_H$, and
$n\mcdot q/m_B = 1-u_H$, we can use the above results to give the final
coefficients for our basis of operators with the minimal number of Dirac
structures, namely
\begin{align} \label{Cresult}
  {\cal C}_9 &=  C_9^{\rm mix}(\mu_0) \bigg\{ 1 + \frac{\alpha_s(\mu_b)}{\pi}
  \bigg[ \omega_a^V(s,\mu_b) + \frac12\, \omega_b^V(s)  +
    \frac{ (1\!-\! u_H)^2\, \omega_c^V(s)}{(1\!-\!
      u_H)^2 \!-\! y_H} \bigg]\bigg\} \nn\\[3pt]
   &\ \ + C_7^{\rm mix}(\mu_0) \frac{\alpha_s(\mu_b)}{\pi}  \bigg[
    \frac{2(1\!-\! u_H)[\,\omega_c^T(s)\!-\!\omega_d^T(s)\,]}{(1\!-\!
      u_H)^2 \!-\! y_H} \!-\! \frac{\omega_b^T(s)}{ (1\!-\! u_H)}  \bigg] \,,
  \nn\\[4pt]
  {\cal C}_7 &=  C_7^{\rm mix}(\mu_0)\bigg\{ 1 + \frac{\alpha_s(\mu_b)}{\pi}
  \bigg[ \omega_a^T(s,\mu_b) - \frac12\, \omega_b^T(s) 
      +  \frac{y_H \, \omega_d^T(s) \!-\! (1\!-\! u_H)^2\omega_c^T(s)
      }{(1\!-\!
      u_H)^2 \!-\! y_H} \bigg]\bigg\} 
   \nn\\[3pt]
   &\ \ - C_9^{\rm mix}(\mu_0) \frac{\alpha_s(\mu_b)}{\pi}  \bigg[
    \frac{y_H\,\omega_b^V(s)}{4 (1\!-\! u_H)} +
    \frac{y_H\,(1\!-\! u_H)\omega_c^V(s)}{2[(1\!-\!
      u_H)^2 \!-\! y_H]} \bigg] \,,
  \nn\\[4pt]
 {\cal C}_{10a} &= C_{10} \Big\{ 1 +
   \frac{\alpha_s(\mu_b)}{\pi}\: \omega_a^V(s,\mu_b) \Big\}  \,, \nn \\[4pt]
  {\cal C}_{10b} &=  C_{10} \: \frac{\alpha_s(\mu_b)}{\pi}\: 
   \Big[ \omega_{b}^V(s) 
   +  \frac{2(1\!-\!u_H)^2 }{(1\!-\!u_H)^2 \!- y_H} \: \omega_{c}^V(s) \Big]
  \,,
\end{align}
where the terms have the structure of a sum over products
$B_1(\mu_0)B_2(\mu_b)$, as desired.

In using the results in Eq.~(\ref{Cresult}) one can choose to work to different
orders in the $\mu_0$- and $\mu_b$-dependent terms, as shown in Eq.~(\ref{mu2}).
For the $\mu_0$ dependence, $C_9^{\rm mix}(\mu_0)$ and $C_7^{\rm mix}(\mu_0)$
include terms from matching at $m_W$ and running to $m_b$, as well as matching
contributions at $m_b$ that cancel the $\mu_0$ dependence from the other pieces.
Thus, these coefficients have only a small residual $\mu_0$ dependence, which is
canceled at higher orders, just as in the local OPE.  The ${\cal C}_i$
coefficients depend on $\mu_b$, both through $\alpha_s(\mu_b)$ and through
explicit $\mu_b$ dependence in $\omega_a^T$ and $\omega_a^V$. The $\ln\mu_b$
dependence in $\omega_a^{V}$ and $\omega_a^{T}$ is identical, as expected from
the known independence of the anomalous dimension on the Dirac structure in
SCET.  The $\mu_b$ dependence in ${\cal C}_i(\mu_b,\mu_0)$ is universal, and
will cancel against the universal $\mu_b$ dependence in the jet and shape
functions, which they multiply in the decay rates. We consider the
phenomenological organization of the perturbative series for $\mu_0$ and $\mu_b$
terms in turn.

First consider the $\mu_0$ terms. Because of mixing, the sizes of contributions
to $C_9^{\rm NDR}$ are comparable at LL and NLL orders~\cite{misiak93,bm95}, so
a reasonable first approximation is to take the NLL result (just as for the 
local OPE decay rate). This entails dropping the ${\cal O}(\alpha_s)$ matching
corrections $C_9^{\rm mix(1)}$ and $C_7^{\rm mix(1)}$, and running $C_9$ at NLL
order with $C_7$ at LL order. As an improved approximation, we would then adopt
the operationally well-defined NNLL approach~\cite{aagw02} of running both $C_9$
and $C_7$ to NLL order and keeping the ${\cal O}(\alpha_s)$ matching corrections
at $m_b$.\footnote{We assume that matching at the high scale, $m_W$, is always
  done at the order appropriate to the running of $U_W(\mu_W,\mu_0)$ in
  Eq.~(\ref{mu2}).}

Below $m_b$ there are Sudakov logarithms. For the $\mu_b$ dependence, the
RG evolution in SCET sums these double-logarithmic series. As a first
approximation we could take the LL and NLL running in $U_H(\mu_b,\mu_i)$ and
$U_S(\mu_i,\mu_\Lambda)$ in Eq.~(\ref{mu2}), while using tree-level matching for
$B_2(\mu_b)$ and ${\cal J}(\mu_i)$. This is consistent because the NLL running
is equivalent to LL running in a single-log resummation. As a second
approximation we could then take NNLL running in both terms and include one-loop
matching for both $B_2(\mu_b)$ and ${\cal J}(\mu_i)$. However since the scales
$m_b^2 \gg m_b\Lambda\gg 1\,{\rm GeV}^2$ are not as well separated as $m_W^2\gg
m_b^2$, we could instead consider the second approximation to include the
one-loop matching for $B_2(\mu_b)$ and ${\cal J}(\mu_i)$ with NLL running, but
without including the full NNLL running (for which parts remain unknown). 

Our procedure for split matching above was based on the non-renormalization of
${\cal O}_{10}$ in QCD. It can also be thought of as matching in two
steps. First one matches at $\mu_0$ on to the scale-invariant operators
\begin{align} \label{Jmix}
  J^{(0)} &= C_9^{\rm mix}\, (\bar s P_R \gamma^\mu b) 
    (\bar\ell \gamma_\mu \ell) +
  C_{10}\: (\bar s P_R \gamma^\mu  b) (\bar\ell \gamma_\mu \gamma_5 \ell)
  \nn\\
  & - C_7^{\rm mix}\, \frac{2 m_B q_\tau}{q^2}
  \big[(\bar s P_R i \sigma^{\mu\tau} b)({\mu=m_b})\big]
  (\bar\ell \gamma_\mu \ell) ,
\end{align}
to determine the coefficients $C_{7,9}^{\rm mix}$. These coefficients are
$\mu_0$ independent at the order in perturbation theory to which the matching is
done.  Secondly, the operators in Eq.~(\ref{Jmix}) are matched on to the SCET
currents in Eq.~(\ref{JscetLO}) at the scale $\mu_b$ to determine the
coefficients ${\cal C}_{7}$, ${\cal C}_9$, ${\cal C}_{10a,b}$. In
Eq.~(\ref{Jmix}) the operators for $C_9^{\rm mix}$ and $C_{10}$ are conserved,
but the tensor current has an anomalous dimension, and so we take $\mu=m_b$ as a
reference point for matching on to a scale-invariant operator. This choice
corresponds to the $\ln m_b$ factor in Eq.~(\ref{C7mix}) for $C_7^{\rm mix}$. A
different choice will affect the division of $\alpha_s(\mu_0)$ or
$\alpha_s(\mu_b)$ terms. Note that Eq.~(\ref{Jmix}) should be thought of only as
an auxiliary step to facilitate the split matching; there is no sense in which
the running of the tensor current is relevant by itself.  In general the
split-matching procedure could be carried out in a manner that gives different
constant terms at a given order, but any such ambiguity will cancel order by
order in ${\cal C}_{7}$ and ${\cal C}_{9}$ (and explicitly if $\mu_0=\mu_b$).
  
Finally, note that our $\omega_a$ differs from the result for
$\omega^{\rm OPE}$ identified in Ref.~\cite{misiak93} for the partonic
semileptonic decay rate when using the local OPE,
\begin{align}
 \omega^{\rm OPE}_{\rm semi} &= -\frac{1}{3} \bigg[ 2\!\ln(s) \ln(1\!-\!s)
  + 4 {\rm Li}_2(s)
  + \ln(1\!-\! s) \Big( \frac{5 \!+\! 4 s}{1\!+\!2s}\Big)
  +\frac{2s(1\!+\! s)(1\!-\! 2s)}{(1\! -\! s)^2 (1\! + \! 2s)} \, \ln(s)
  \nn\\
  & \qquad\quad - \frac{(5\!+\! 9s \!- \! 6s^2)}{2(1\! -\! s)(1\!+\! 2s)}
   + \frac{2\pi^2}{3}  \bigg] \,.
\end{align}
Here $\omega_{\rm semi}^{\rm OPE}$ contains both vertex and bremsstrahlung
contributions evaluated in the full theory. Grouping these contributions with
the Wilson coefficient for ${\cal O}_9$ gives
\begin{align}
 C_9^{\rm local}(\mu) = C_9^{\rm mix}(\mu) + P_0^{\rm NDR}(\mu)\: 
    \frac{\alpha_s(\mu)}{\pi} \: \omega^{\rm OPE}_{\rm semi} \,,
\end{align}  
which is $\tilde C_9^{\rm eff}$ in the notation in Ref.~\cite{bm95}.  At LO, the
restricted phase space in the shape function region causes bremsstrahlung to
contribute only to the jet and shape functions, and not at the scale $\mu\simeq
m_b$. The shape function and jet function also modify the contributions from the
vertex graphs. Thus, instead of $\omega^{\rm OPE}_{\rm semi}$ the final results
in the shape function region are given by our $\omega_i^V$ and $\omega_i^T$ 
factors appearing in $C_{9i}$ and $C_{7i}$. 
Consequently, the main difference is in the
terms we match at $\mu=\mu_b$, while the terms matched at $\mu=\mu_0$ that
appear in $C_9^{\rm mix}$ and $C_7^{\rm mix}$ are identical to terms appearing
in the local OPE analysis.

\subsection{RG Evolution Between $\mu_b$ and $\mu_i$} \label{Run1}

The running of the Wilson coefficients in SCET from the scale $\mu_b^2\sim
m_b^2$ to $\mu_i^2\sim m_b \Lambda_{\rm QCD}$ involves double Sudakov logarithms
and was derived in Refs.~\cite{bfl01,bfps01} at NLL order. The SCET running is
independent of the Dirac structure of the currents, which is a reflection of the
spin symmetry structure of the current.  We briefly outline a short argument for
why this is true to all orders in perturbation theory.  The leading-order
currents in SCET have the structure
\begin{align}
  J = (\bar\xi_n W)_p \Gamma (Y^\dagger h_v) \,,
\end{align}
and we wish to see that their anomalous dimension is independent of $\Gamma$.
The anomalous dimensions are computed from the UV structure of SCET loop
diagrams, with the Lagrangians in Eq.~(\ref{LLO}). Soft gluon loops involve
contractions between the Wilson line $Y^\dagger$ and the $h_v$ and do not change
the Dirac structure. Next consider the collinear loops. The attachment of a
gluon from the Wilson line $W$ to the collinear quark gives a factor of a
projection matrix, which can be pushed through $\gamma_\perp$'s to give
$\bar\xi_n \bnslash\nslash/4 = \bar\xi_n$. Thus it does not modify the Dirac
structure, so only insertions from the \hbox{$i \Dslash^\perp_c {1}/({i\bn\mcdot
    D_c}) i \Dslash^\perp_c$} term are of concern. These terms give structures
of the form $\bar u_n^{(u)} \gamma_\perp^{\mu_1} \gamma_\perp^{\mu_2}\cdots
\gamma_\perp^{\mu_{2k}} \Gamma u_v^{(b)}$, where all $\mu_i$ indices are
contracted with each other. Using $\{\gamma_\perp^\mu,\gamma_\perp^\nu\} =
2g_\perp^{\mu\nu}$ and $\gamma_\perp^\mu \gamma^\perp_\mu =d-2$, we can reduce
this product to terms with zero $\gamma_\perp$'s since all vector indices are
contracted. Hence all diagrams reduce to having the Dirac structure that was
present at tree level, $\bar u_n^{(u)} \Gamma u_v^{(b)}$.

Thus, all the LO coefficients obey the same homogeneous anomalous
dimension equation,
\begin{align}  \label{rge2}
   \mu \frac{d}{d\mu} {\cal C}_i(\mu) &= 
  \Big[ - \Gamma_{\rm cusp}(\alpha_s) \ln\Big(\frac{\mu}{\bnP}\Big)
   + \tilde \gamma(\alpha_s) \Big]  {\cal C}_i(\mu) \\
  &= 
  \bigg[ - \Gamma_{\rm cusp}(\alpha_s) \ln\Big(\frac{\mu}{\mu_b}\Big)
   + \Big\{\tilde \gamma(\alpha_s)
   + \Gamma_{\rm cusp}(\alpha_s) \ln\Big(\frac{\bn\mcdot p}{\mu_b}\Big) \Big\}
   \bigg]  {\cal C}_i(\mu) \,.\nn
\end{align}
This must be integrated together with the beta function $\beta=\mu d/d\mu\:
\alpha_s(\mu)$ to solve for $U_H$ in
\begin{align} \label{CUHC}
 {\cal C}_i(\mu_i) &= \sqrt{U_H(\mu_i,\mu_b)}\: {\cal C}_i(\mu_b)\,.
\end{align}  
In the second line of Eq.~(\ref{rge2}) we used the fact that $\bnP$ gives the
total partonic $\bn\mcdot p$ momentum of the jet $X_s$ in the $B\to
X_s\ell^+\ell^-$ matrix element, and we introduced artificial dependence on the
matching scale $\mu_b$ in order to make the $\bn\mcdot p$ dependence appear in
a small logarithm. Here $\bn\mcdot p = m_b - \bn\mcdot q$. We write
\begin{align}
  \Gamma^{\rm cusp} &= 
   \sum_{n=0}^\infty \Gamma_n^{\rm cusp} \Big(\frac{\alpha_s}{4\pi}\Big)^{n+1} \,,
  & \tilde \gamma &= 
   \sum_{n=0}^\infty \tilde\gamma_n \Big(\frac{\alpha_s}{4\pi}\Big)^{n+1} \,,
  &\beta &  =  
    - 2 \alpha_s \sum_{n=0}^\infty \beta_n \Big(\frac{\alpha_s}{4\pi}\Big)^{n+1} \,.
\end{align}
At NLL order we need $\beta_0=11 C_A/3- 2n_f/3$, $\beta_1 = 34 C_A^2/3 -10 C_A
n_f/3 - 2 C_F n_f$ and
\begin{align}
 \Gamma_0^{\rm cusp} &= 4 C_F \,,
 &\Gamma_1^{\rm cusp} &= 8 C_F B = 8 C_F \Big[ C_A \Big( \frac{67}{18}\!-\!\frac{\pi^2}{6}\Big) 
      \!-\! \frac{5}{9}n_f \Big] \,,
 &\tilde \gamma_0 &= -5 C_F \,,
\end{align}
where $C_A=3$ and $C_F=4/3$ for SU(3). For the number of active flavors we take
$n_f=4$ since we're running below $m_b$.  The cusp anomalous dimension
$\Gamma_1^{\rm cusp}$ was computed in Ref.~\cite{KRKK}, and the result for
$\Gamma_2^{\rm cusp}$ was recently found in Ref.~\cite{Moch:2004pa}.  RG
evolution in SCET at NNLL order has been considered in
Refs.~\cite{neubert,Lange:2005qn}. For the NNLL result one needs $\Gamma_2^{\rm
  cusp}$, $\tilde \gamma_1$, and $\beta_2$. For $\tilde \gamma_1$ an independent
calculation does not exist, but a conjecture for its value was given in
Ref.~\cite{neubert} based on the structure of the three-loop splitting
function~\cite{Moch:2004pa}. For the sake of clarity we stick to NLL order
here.  The result is
\begin{eqnarray} \label{sln1}
   U_H(\mu_i,\mu_b)= \exp\Bigg[\frac{2g_0(r_1)}{\alpha_s(\mu_b)}
  + 2g_1(r_1,\bn\mcdot p) \bigg]  \,,
\end{eqnarray}
where the independent variable is $\mu_i$ and
\begin{eqnarray} \label{r1}
  r_1(\mu_i) = \frac{\alpha_s(\mu_i)}{\alpha_s(\mu_b)} = \frac{2\pi}{2\pi+ {\beta_0}
  \alpha_s(\mu_b)\ln(\mu_i/\mu_b) }\,,
\end{eqnarray}
with
\begin{eqnarray} \label{LONLOC}
 g_0(r_1) & = & -\frac{4\pi C_F}{ \beta_0^2}
 \:\Big[ \frac{1}{r_1} -1 + \ln r_1 \Big] \,,
\\
 g_1(r_1, \bn\!\cdot\!  p)
 &=& -\frac{ C_F\beta_1}{\beta_0^3} \Big[ 1 -r_1 + r_1\ln r_1 
  -\frac12 \ln^2 r_1 \Big]\nn \\
 && + \frac{C_F}{\beta_0} \Big[ \frac52 - 2 \ln\Big(\frac{\bn\!\cdot\! p}{\mu_b}\Big) \Big] 
  \ln r_1 - \frac{2 C_F B}{\beta_0^2}  \Big[ r_1 -1- \ln r_1 \Big] \,. \nn
\end{eqnarray}
This is the form for the universal running of the LO SCET currents found in
Ref.~\cite{bfps01}.  Switching to $\alpha_s(\mu_i)$ as the independent variable,
with $r_1=\alpha_s(\mu_i)/\alpha_s(\mu_b)$, gives
\begin{eqnarray} \label{sln1a}
   U_H(\mu_i,\mu_b)= \Big( \frac{\bn\mcdot p}{\mu_b}\Big)^{\!-\frac{4C_F}
   {\beta_0}\,\ln r_1} 
   \exp\Bigg[\frac{2g_0(r_1)}{\alpha_s(\mu_b)}
  + 2\tilde g_1(r_1) \bigg]  \,,
\end{eqnarray}
where $g_0(r_1)$ is as in Eq.~(\ref{LONLOC}) and 
\begin{align}
  \tilde g_1(r_1) &=\frac{ C_F\beta_1}{2\beta_0^3}\ln^2 r_1  
  + \frac{5C_F}{2\beta_0}
  \ln r_1 + \frac{C_F}{\beta_0^3}\big({2 B\beta_0}-\beta_1\big)\big( 1 -r_1 + \ln r_1 
  \big)  \,. 
\end{align}
This form of the evolution with $\alpha_s(\mu)$ as the variable was used in
Ref.~\cite{blnp04}, and is also the one we adopt here.  The decay rate is
computed from a time-ordered product of currents and so at the intermediate
scale $\mu_i^2\sim m_b\Lambda$ will involve products
\begin{align}
 {\cal C}_i(\mu_i,\mu_0)\, {\cal C}_j(\mu_i,\mu_0) = U_H(\mu_i,\mu_b)\,
  {\cal C}_i(\mu_b,\mu_0)\, {\cal C}_j(\mu_b,\mu_0)  \,,
\end{align}
explaining why we used a notation with $\sqrt{U_H}$ in Eq.~(\ref{CUHC}).

\subsection{Hadronic Tensor and Decay Rates} \label{decay}

In the last two sections we constructed the required basis of SCET current
operators with matching at $\mu_0^2\sim \mu_b^2\sim m_b^2$ and evolution to
$\mu_i^2\sim m_b\Lambda$.  At the scale $\mu_i$ we take time-ordered products of
the SCET currents and compute the decay rates using the optical theorem.  In
this section we discuss the tensor decomposition of the time-ordered products
and results for differential decay rates.

In order to simplify the computation of decay rates it is useful to write the
sum of hadronic operators as a sum of left-handed and right-handed terms since
for massless leptons we have only LL or RR contributions~\cite{ahhm97}.  Doing
this for our current, we have
\begin{eqnarray}
  {\cal J}^{(0)}_{\ell\ell} & = &
   \big[ {\cal C}_{9} - {\cal C}_{10a} \big]
          \big( \bar\chi_n \,   \gamma_\mu P_L\, {\cal H}_v \big)
          \big( \bar{\ell} \, \gamma^\mu  P_L \, \ell \big)
          +  \big[ {\cal C}_{9} + {\cal C}_{10a} \big]
          \big( \bar\chi_n \, \gamma_\mu P_L\, {\cal H}_v \big)
          \big( \bar{\ell} \, \gamma^\mu  P_R \, \ell \big)
    \nn \\[4pt]
  & &
     + {\cal C}_{10b} \big( \bar\chi_n \, v_\mu P_R\, {\cal H}_v \big)
      \big( \bar{\ell} \, \gamma^\mu \gamma_5 \, \ell \big)
  -  {\cal C}_{7}\, \frac{2 m_B q^\tau}{q^2} \big( \bar\chi_n \, i\sigma_{\mu\tau} 
     \, {\cal H}_v \big)
          \big( \bar{l} \, \gamma^\mu \, l \big)
                 \nn \\ 
 & = &
  \left( J_{L\mu} \, L_L^\mu 
  + J_{R\mu} \, L_R^\mu \right) \, ,
\end{eqnarray}
where
\begin{eqnarray}
  L_{L}^\mu & = & \bar{\ell} \, \gamma^\mu \, P_{L} \, \ell \, ,
  \qquad 
   L_{R}^\mu  =  \bar{\ell} \, \gamma^\mu \, P_{R} \, \ell \, , \\
  J_{L(R)}^\mu & = & \bar\chi_n \, P_R \,  
   \Big[  \big( {\cal C}_{9} \mp {\cal C}_{10a} \big) \gamma^\mu 
   + {\cal C}_{7} \, \frac{2 m_B \gamma^\mu\, {\slash\!\!\! q}}{{q}^2} 
   \mp {\cal C}_{10b}\, v^\mu 
   \Big] {\cal H}_v \nn \\  
  &\equiv & \bar\chi_n \, \Gamma_{L(R)}^\mu \, {\cal H}_v \, . \nn
\end{eqnarray}

Thus, the inclusive decay rate for $\bar B\to X_s \ell^+ \ell^-$ 
is proportional to $(W^L_{\mu\nu}L_L^{\mu\nu} + W^R_{\mu\nu}L_R^{\mu\nu})$,
where the leptonic parts $L_{L(R)}^{\mu\nu}$ and hadronic parts 
$W_{L(R)}^{\mu\nu}$ are given by
\begin{eqnarray} 
L_{L(R)}^{\mu\nu} & = & \sum_{\mbox{\small spin}}
  \left[ \bar{l}_{L(R)}(p_+) \, \gamma^\mu \, l_{L(R)}(p_-) \right]
  \left[ \bar{l}_{L(R)}(p_-) \, \gamma^\nu \, l_{L(R)}(p_+) \right]
                \nn \\
 & = & 2 \left[ p_+^\mu \, p_-^\nu + p_-^\mu \, p_+^\nu
   - g^{\mu\nu}\, p_+ \!\cdot \! p_- \mp i \epsilon^{\mu\nu\alpha\beta} \,
                      p_{+\alpha} \, p_{-\beta} \right] , 
   \label{Lmunu}
\end{eqnarray}
and
\begin{eqnarray} \label{defnW}
  W_{\mu\nu}^{L(R)} &=& \frac{1}{2m_B} \sum_X (2\pi)^3 \delta^4(p_B-q-p_X)
  \langle \bar B | J_\mu^{L(R)\dagger} | X \rangle \langle X | J_\nu^{L(R)} 
  | \bar B \rangle \\
  &=& - g_{\mu\nu} W_1^{L(R)} + v_\mu v_\nu W_2^{L(R)} 
  + i\epsilon_{\mu\nu\alpha\beta} v^\alpha q^\beta W_3^{L(R)} 
  + q_\mu q_\nu W_4^{L(R)} + (v_\mu q_\nu + v_\nu q_\mu) W_5^{L(R)}
  \,. \nn
\end{eqnarray}
Here, we use 
relativistic normalization for the $|\bar B\rangle$ states.  
For convenience, we define projection tensors 
$P_i^{\mu\nu}$ so that
\begin{eqnarray}
   W_i^{L(R)} = P_i^{\mu\nu}\: W_{\mu\nu}^{L(R)} \,.
\end{eqnarray}
They are
\begin{eqnarray} \label{ProjW}
  P_1^{\mu\nu} &=& -\frac{1}{2} g^{\mu\nu}
  + \frac{q^2\, v^{\mu}v^\nu + q^\mu q^\nu  - v\mcdot q (v^\mu q^\nu+v^\nu
    q^\mu)} {2 [q^2-(v\mcdot q)^2]}\,,\\[5pt]
  P_2^{\mu\nu} &=& \frac{3 q^2\, P_1^{\mu\nu} + q^2 g^{\mu\nu} -q^\mu q^\nu}
    { [q^2-(v\mcdot q)^2]} \,,\qquad
  P_3^{\mu\nu} = \frac{-i \epsilon^{\mu\nu\alpha\beta} q_\alpha v_\beta}
    { 2[q^2-(v\mcdot q)^2]} \,,\nn \\[5pt]
  P_4^{\mu\nu} &=& \frac{g^{\mu\nu} - v^\mu v^\nu + 3 P_1^{\mu\nu}}
    { [q^2-(v\mcdot q)^2]} \,, \qquad
  P_5^{\mu\nu} = \frac{g^{\mu\nu} +  4 P_1^{\mu\nu}-P_2^{\mu\nu} - q^2
    P_4^{\mu\nu}} {2 v\mcdot q} \,.\nn
\end{eqnarray}
The optical theorem relates $W_{\mu\nu}^{L(R)}$ to the forward-scattering
amplitude defined as
\begin{eqnarray} \label{T}
  T_{\mu\nu}^{L} &=& \frac{-i}{2 m_B} \int\!\! d^4x\, e^{-iq\cdot x}\,
  \langle \bar B|  T J_\mu^{L\dagger}(x) J_\nu^{L}(0)
  |\bar B\rangle \\
   &=& - g_{\mu\nu} T_1^{L} + v_\mu v_\nu T_2^{L}
   + i\epsilon_{\mu\nu\alpha\beta} v^\alpha q^\beta T_3^{L}
   + q_\mu q_\nu T_4^{L} + (v_\mu q_\nu + v_\nu q_\mu) T_5^{L}
  \,,\nn
\end{eqnarray}
with an analogous definition for $T_{\mu\nu}^{R}$, giving
\begin{eqnarray} \label{WT}
  W_i^{L} = -\frac{1}{\pi} \:{\rm Im}\, T_i^{L} \,,\qquad
   W_i^{R} = -\frac{1}{\pi} \:{\rm Im}\, T_i^{R} \,.
\end{eqnarray}

Contracting the lepton tensor $L_{L(R)}^{\mu\nu}$ with $W_{L(R)}^{\mu\nu}$ 
and neglecting the mass of the leptons give the differential decay rate
\begin{align} \label{dGamma}
 \frac{d^3\Gamma}{dq^2 dE_-  dE_+}  
    &= \Gamma_0\: \frac{96}{m_B^5} \Big[ q^2 W_1 + 
    (2 E_- E_+ -q^2/2) W_2  + q^2 (E_--E_+) W_3 \Big]
    \theta(4 E_-E_+ - q^2) \,, 
\end{align}
where $E_\pm = v\cdot p_\pm$, $W_1=W_1^L+W_1^R$, $W_2=W_2^L+W_2^R$,
$W_3=W_3^L-W_3^R$ and the normalization factor is
\begin{eqnarray}
 \Gamma_0 = \frac{G_F^2\, m_B^5}{192\pi^3}\,
  \frac{\alpha^2}{16\pi^2}\, |V_{tb} V_{ts}^*|^2\,.
\end{eqnarray}
The $W_i$ are functions of $q^2$ and $v\cdot q = v \cdot (p_+ + p_-)$. Another
quantity of interest is the forward-backward asymmetry in the variable
\begin{align}
  \cos\theta = \frac{v\cdot p_- - v\cdot p_+}{\sqrt{(v\cdot q)^2 -q^2}} ,
\end{align}
where $\theta$ is the angle between the $B$ and $\ell^+$ in the CM frame of the
$\ell^+\ell^-$ pair:
\begin{align}
 \frac{d^2 A_{\rm FB}}{dv\mcdot q\: dq^2} \equiv \int_{-1}^1\!\! d(\cos\theta)\:
    \frac{{\rm sign}(\cos\theta)}{\Gamma_0} 
   \frac{d^3 \Gamma}{dv\mcdot q\: dq^2\: d\cos\theta}
  = \frac{48\, q^2}{m_B^5} \big[ (v\cdot q)^2 - q^2 \big]\: W_3 \,.
\end{align} 

In terms of the dimensionless variables
\begin{eqnarray} \label{dimless}
  x_H = \frac{2 E_{\ell^-}}{m_B} \,, \qquad
  \overline y_H = \frac{\bn\mcdot p_X}{m_B} \,,\qquad 
  u_H = \frac{n\mcdot p_X}{m_B} \,,
\end{eqnarray}
the triply differential decay rate is
\begin{align} \label{dGamma3u}
 \frac{1}{\Gamma_0} \frac{d^3\Gamma}{ dx_H\, d\overline y_H\, du_H}  
    &= {24m_B} (\overline y_H\!-\! u_H)
   \Big\{(1\!-\! u_H)(1\!-\! \overline y_H) W_1 + 
    \frac12 (1\!-\!x_H\!-\!u_H)(x_H\!+\!\overline y_H\!-\!1) W_2 \nn\\
 & \hspace{0.5cm}
  + \frac{m_B}{2}\,(1\!-\! u_H)(1\!-\! \overline y_H)
     \big(2x_H\!+\! u_H \!+\!\overline y_H\!-\! 2\big) W_3 \Big\}
    \,,
\end{align}
where $W_i=W_i(u_H,\overline y_H)$.  For a strict SCET expansion we want
${n\mcdot p_X} \ll { \bn\mcdot p_X} $ i.e.\ $u_H \ll \overline y_H$. However, it
is useful to keep the full dependence on the phase-space prefactors rather than
expanding them, because it is then simpler to make contact with the total rate in the
local OPE, as emphasized recently in Refs.~\cite{frank,bjorn}, and so we keep
these factors here. We shall also keep the formally subleading kinematic
prefactors in our hard functions rather than expanding them as we did in
Ref.~\cite{klis04}. Other variables of interest include the dilepton and
hadronic invariant masses,
\begin{eqnarray}
  \quad y_H = \frac{q^2}{m_B^2}\,, \qquad
  s_H = \frac{m_X^2}{m_B^2} \,,
\end{eqnarray}
where
\begin{eqnarray} 
   s_H = u_H \overline y_H  \,,\qquad
    y_H = (1-u_H)(1-\overline y_H) \,,
   \label{eq:sy}
\end{eqnarray}
so that [$\overline y_H \ge u_H$]
\begin{eqnarray}
 \{ \overline y_H , u_H\} &=& \frac{1}{2} \Big[ 1-y_H+s_H
    \pm \sqrt{(1-y_H+s_H)^2-4 s_H}\,\Big] \,.
   \label{eq:yu}
\end{eqnarray}
A few interesting doubly differential spectra  are
\begin{align}  \label{dGamma2up}
 \frac{1}{\Gamma_0} \frac{d^2\Gamma}{  d\overline y_H\, du_H}
    &= {24m_B} (\overline y_H\!-\! u_H)^2
   \Big\{(1\!-\! u_H)(1\!-\! \overline y_H)  W_1 +
    \frac{1}{12} (\overline y_H\!-\! u_H)^2 W_2
  \Big\}
     \,,\\[4pt]
\frac{1}{\Gamma_0} \frac{d^2\Gamma}{  dy_H\, ds_H}
    &= {2m_B} \sqrt{(1\!-\!y_H\!+\! s_H)^2\!-\! 4\, s_H}\:
   \Big\{ 12 y_H W_1  
   +\big[(1\!-\!y_H\!+\! s_H)^2 \!-\! 4 s_H\big] W_2 \Big\}
 \,, \nn\\[4pt]
 \frac{1}{\Gamma_0} \frac{d^2\Gamma}{  dy_H\, du_H} &= 
   \frac{2 m_B}{(1\!-\! u_H)^3} \big[ (1\!-\! u_H)^2\!-\!y_H \big]^2
   \Big\{ 12 y_H W_1 + \Big[ \frac{(1\!-\! u_H)^2\!-\! y_H}{(1\!-\! u_H)}
   \Big]^2 W_2 \Big\}
  \,,\nn\\[4pt]
 \frac{1}{\Gamma_0} \frac{d^2\Gamma}{  ds_H\, du_H} &= 
   \frac{2 m_B (s_H\!-\! u_H^2)^2}{u_H^5} 
   \Big\{ 12 u_H (1\!-\!u_H)(u_H\!-\! s_H)  W_1 + (s_H\!-\!u_H^2)^2\: W_2 \Big\}
  \,.\nn
\end{align}
For doubly differential forward-backward asymmetries we find
\begin{align} \label{dAFB2up}
\frac{d^2 A_{\rm FB}}{d\overline y_H\: d u_H} 
  &=  6 m_B^2\, (\overline y_H\!-\!u_H)^3 (1\!-\!u_H)(1\!-\!\overline y_H)\: W_3
  \,,\\[4pt]
  \frac{d^2 A_{\rm FB}}{  dy_H\, ds_H}
  &= 6 m_B^2\, y_H \big[ (1\!-\! y_H \!+\! s_H)^2 -4 s_H\big] W_3
   \,,\nn\\[4pt]
  \frac{d^2 A_{\rm FB}}{  dy_H\, du_H} 
  &= 6 m_B^2\, \frac{y_H\,[(1\! -\! u_H)^2\!-\! y_H]^3 }{(1\! -\! u_H)^4}\: W_3 
  \,, \nn\\[4pt]
  \frac{d^2 A_{\rm FB}}{  ds_H\, du_H} 
  &=6 m_B^2\, \frac{(s_H\!-\! u_H^2)^3 (u_H\!-\!s_H) (1\!-\!u_H)}{u_H^5}\: W_3 
  \,.\nn
\end{align}

\subsection{LO Matrix Elements in SCET} \label{SCET}

\begin{figure}[t!]
 \centerline{
  \mbox{\epsfysize=3.truecm \hbox{\epsfbox{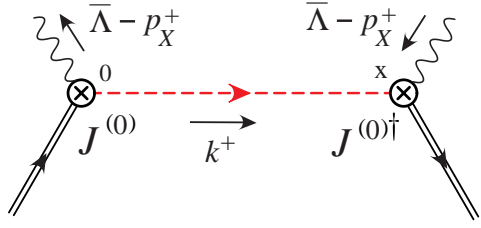}} }
  }
 {\caption[1]{Time-ordered product for the leading-order factorization theorem.}
\label{fig:LO_TO_prod} }
 \vskip -0.1cm
\end{figure}

At lowest order in the $\Lambda/m_b$ expansion, the only time-ordered product
consists of two lowest-order currents ${\cal J}_{\ell\ell}^{(0)}$ as shown in
Fig.~\ref{fig:LO_TO_prod}. The factorization of hard contributions into the SCET
Wilson coefficients and the decoupling of soft and collinear gluons at lowest
order are identical to the steps for $B\to X_s\gamma$ and $B\to X_u\ell\bar\nu$,
and directly give the factorization theorem for these time-ordered
products~\cite{bps02}. The SCET result agrees with the factorization theorem of
Korchemsky and Sterman~\cite{ks94}. However, the structure of
$\alpha_s(\sqrt{m_b\Lambda})$ and $\alpha_s(m_b)$ corrections differs from
the parton-model rate, as mentioned in Refs.~\cite{bm04,blnp04}. Beyond lowest
order in $\alpha_s(m_b)$ the kinematic dependences also differ, as mentioned in
Ref.~\cite{klis04}. For $B\to X_u\ell\bar\nu$, the final triply differential rate
with perturbative corrections at ${\cal O}(\alpha_s)$ can be found in
Refs.~\cite{bm04,blnp04}.

The factorization and use of the optical theorem is carried out at the scale
$\mu=\mu_i$, and we expand $W_i = W_i^{(0)}+ W_i^{(2)} + \ldots$ in powers of
$\lambda = (\Lambda_{\rm QCD}/m_b)^{1/2}$ (with no linear term). For $B\to
X_s\ell^+\ell^-$ we have bilinear hadronic current operators in SCET in
Eq.~(\ref{JscetLO}) and so, as is the case for $B\to X_u\ell\bar\nu$, we find
\begin{eqnarray}
W_{i}^{(0)} & = &
  h_i(p_X^+,p_X^-,\mu_i)
  \int_{0}^{p_X^+} \! \! dk^+\:
   {\cal J}^{(0)}(p^-\,, k^+,\mu_i )\:
   f^{(0)}(k^+  \!+\!\overline\Lambda\!-\! p_X^+,\mu_i)
 \,.
   \label{WiLO}
\end{eqnarray}
This result is important, since it states that the same shape function $f^{(0)}$
appears in $B\to X_s\ell^+\ell^-$ as appears in $B\to X_s\gamma$ and $B\to
X_u\ell\bar\nu$.  This formula relies on the power counting $s\sim y_H\sim
\lambda^0$ that we adopted (and would not be true for the counting $s\sim
\lambda^2$ discussed in Appendix~\ref{App:q2jet}).  At tree level the structure
of this factorization theorem is illustrated by Fig.~\ref{fig:LO_TO_prod}. The
hard coefficients here are
\begin{align} \label{LOTr}
 h_1(p_X^+,p_X^-,\mu_i) &= 
  \frac14 {\rm Tr} \Big[ {P_v}\, \overline \Gamma_{\mu}^{L} \,
  {\nslash}\, \Gamma_{\nu}^{L} \Big] P_1^{\mu\nu} +
  \frac14
  {\rm Tr} \Big[ {P_v}\, \overline \Gamma_{\mu}^{R} \,
  {\nslash}\, \Gamma_{\nu}^{R} \Big] P_1^{\mu\nu} \,, \nn\\
 h_2(p_X^+,p_X^-,\mu_i) &= \frac14 {\rm Tr} \Big[ {P_v}\, \overline \Gamma_{\mu}^{L} \,
  {\nslash}\, \Gamma_{\nu}^{L} \Big] P_2^{\mu\nu} +
  \frac14
  {\rm Tr} \Big[ {P_v}\, \overline \Gamma_{\mu}^{R} \,
  {\nslash}\, \Gamma_{\nu}^{R} \Big] P_2^{\mu\nu} \,, \nn\\
h_3(p_X^+,p_X^-,\mu_i) &= 
  \frac14 {\rm Tr} \Big[ {P_v}\, \overline \Gamma_{\mu}^{L} \,
  {\nslash}\, \Gamma_{\nu}^{L} \Big] P_3^{\mu\nu} -
  \frac14
  {\rm Tr} \Big[ {P_v}\, \overline \Gamma_{\mu}^{R} \,
  {\nslash}\, \Gamma_{\nu}^{R} \Big] P_3^{\mu\nu}\,,
\end{align}
with $P_v = (1+\vslash)/2 $ and $\overline\Gamma = \gamma^0 \Gamma^\dagger
\gamma^0$.  In Eq.~(\ref{WiLO}) we have the same leading-order shape function as
in $B\to X_s\gamma$ and $B\to X_u\ell\bar\nu$, namely
\begin{eqnarray} \label{f0}
  f^{(0)}(\ell^+,\mu_i) &=&
  \frac{1}{2} \int\!\! \frac{dx^-}{4\pi} e^{-i x^- \ell^+/2}
   \ \langle \bar B_v | \bar {\cal H}_v(\tilde x) {\cal H}_v(0)
  | \bar B_v \rangle \nn\\
  &=&\frac{1}{2}\: \langle \bar B_v | \bar h_v
  \delta(\ell^+\!-\! in\mcdot D)  h_v | \bar B_v \rangle \,,
\end{eqnarray}
where $\tilde x^\mu = \bn\mcdot x \, n^\mu/2$. This function was first discussed
in Ref.~\cite{Neubert:1993ch}.  The jet function is defined by ${\cal J}^{(0)}
(p^-,k^+)=(-1/\pi)\,{\rm Im}\, {\cal J}_{\omega=p^-}^{(0)}(k^+)$
$\times\theta(p_X^+-k^+)$, where
\begin{align}
  \label{Jmelt}
  &\hspace{-0.9cm}
 i\,\big\langle 0 \big| T \big[ \bar\chi_{n,\omega}(0) \frac{\bnslash}{4N_c}
    \chi_{n,\omega'}(x) \big] \big| 0 \big\rangle 
 = 
    \delta(\omega\!-\!\omega') \delta^2(x_\perp) \delta(x^+) 
    \int\!\! \frac{dk^+}{2\pi}\: e^{-i k^+ x^-/2}\ {\cal J}^{(0)}_\omega(k^+) \,,
\end{align}
and is known at one-loop order~\cite{bm04,blnp04}, namely
\begin{eqnarray} \label{J01loop2}
  {\cal J}^{(0)}(p^-,z\,p_X^+,\mu_i) &=& 
 \frac{1}{p_X^+}
   \bigg\{ \delta(z) \bigg[
   1 + \frac{\alpha_s(\mu_i)C_F}{4\pi} \Big( 2 \ln^2\frac{p^-p_X^+}{\mu_i^2} -
   3 \ln \frac{p^- p_X^+}{\mu_i^2} + 7-\pi^2 \Big) \bigg] \\
 && +\frac{\alpha_s(\mu_i)C_F}{4\pi} \bigg[ \Big(\frac{4\ln z}{z}\Big)_+ +
  \Big(4 \ln\frac{p^- p_X^+}{\mu_i^2}-3 \Big)\: \frac{1}{(z)_+} \bigg]
  \theta(z) \,  \bigg\}
   \theta(1\!-\!z)\,,\nn
\end{eqnarray}
where $z = k^+/p_X^+$. Despite appearances, only the combination $zp_X^+$
appears in ${\cal J}^{(0)}$ apart from the $\theta(1-z)$. This last
$\theta$-function is induced by the soft function and, when one takes the
imaginary part of the full time-ordered product, affects the complex structure.
Therefore, we include it in our definition of ${\cal J}^{(0)}(p^-,k^+)$.

\subsection{RG Evolution Between $\mu_\Lambda$ and $\mu_i$ } \label{Run}

The function $f^{(0)}$ cannot be computed in perturbation theory and must
therefore be extracted from data. This same function appears at LO in the $B\to
X_s\gamma$, $B\to X_u \ell\bar\nu$ and $B\to X_s\ell^+\ell^-$ decay rates. In
practice, a model for $f^{(0)}$ is written down with a few parameters, which are
fitted to the data. The support of $f^{(0)}(\overline \Lambda-r^+)$ is $-\infty$ to
$\overline\Lambda$ since $r^+ \in [ 0 , \infty)$. It is often convenient to
switch variables to $\hat f^{(0)}(r^+)=f^{(0)}(\overline \Lambda-r^+)$ which has
support from $0$ to $\infty$, although we shall keep using $f^{(0)}$ here. A
typical three-parameter model is~\cite{Leibovich:2002ys,bjorn}
\begin{eqnarray}
  f^{(0)}(\overline \Lambda-r^+,\mu_\Lambda) = \hat f^{(0)}(r^+,\mu_\Lambda) 
  = \frac{a^b \, (r^+)^{b-1}}{\Gamma(b)\,
    L^b} \exp\Big(\frac{-a r^+}{L}\Big) \theta(r^+) \,,
\end{eqnarray}
where $a,b$ are dimensionless and $L\sim \Lambda_{\rm QCD}$. These parameters
can be fitted to the $B\to X_s\gamma$ photon spectrum and the function $f^{(0)}$
can then be used elsewhere.  The most natural scale to fix this model at is
$\mu=\mu_\Lambda \sim 1\,{\rm GeV}$, at which it contains no large logarithms.
The result of evolving the shape function to the intermediate scale is
then~\cite{blnp04}
\begin{eqnarray}
 f^{(0)}(\overline\Lambda\!-\!r^+,\mu_i) & = & e^{V_S(\mu_i,\mu_\Lambda)}\,
 \frac{1}{\Gamma(\eta)} \int_0^{r^+}\!\! d r^{+\prime}\:
 \frac{f^{(0)}(\overline\Lambda\!-\!r^{+\prime},\mu_\Lambda)}
 {\mu_\Lambda^\eta\,(r^+-r^{+\prime})^{1-\eta}} \,.
\end{eqnarray}
(The structure of this result also applies at higher orders in 
RG-improved perturbation theory~\cite{neubert}, and at one-loop order a
similar structure was considered earlier, in Ref.~\cite{Balzereit:1998yf}.) At
NLL order
\begin{eqnarray}
   V_S(\mu_i,\mu_\Lambda)
   &=& \frac{\Gamma_0^{\rm cusp}}{2\beta_0^2} \left[
    \frac{-4\pi}{\alpha_s(\mu_\Lambda)}\,(r_2 \!-\! 1 \!-\! \ln r_2)
    \!+\! \frac{\beta_1}{2\beta_0}\,\ln^2 r_2
    \!+\! \left( \frac{\Gamma_1^{\rm cusp}}{\Gamma_0^{\rm cusp}} 
   \!- \frac{\beta_1}{\beta_0} \right)
    \left( 1\!-\! \frac{1}{r_2} \!-\! \ln r_2 \right) \right] \nonumber\\
   &&\mbox{}- \frac{\Gamma_0^{\rm cusp}}{\beta_0}\,\gamma_E\,\ln r_2
    - \frac{\gamma_0}{\beta_0}\,\ln r_2 \,, \nn\\
  \eta &=& \frac{\Gamma_0^{\rm cusp}}{\beta_0}\: \ln r_2  \,.
\end{eqnarray}
Here, $r_2 = \alpha_s(\mu_\Lambda)/\alpha_s(\mu_i)$, $\Gamma_0^{\rm cusp}$ and
$\Gamma_1^{\rm cusp}$ are the same as in Section~\ref{Run1} and $\gamma_0 = -2
C_F$. 
For numerical integration this can be rewritten in the form
\begin{eqnarray}
 f^{(0)}(\overline\Lambda\!-\!r^+,\mu_i) & = & e^{V_S(\mu_i,\mu_\Lambda)}\,
 \frac{1}{\Gamma(1+\eta)} \left(\frac{r^+}{\mu_\Lambda}\right)^\eta
\int_0^1 \!dt\: f^{(0)}\left(\overline\Lambda\!-\! r^+ (1 - t^{1/\eta}), 
 \mu_\Lambda\right) \,.
\end{eqnarray} 



\section{$B\to X_s\ell^+\ell^-$ Spectra in the Shape Function Region
} \label{results}

\subsection{Triply Differential Spectrum}

At lowest order in the power expansion, Eqs.~(\ref{ProjW}) and (\ref{WiLO}) give
the result
\begin{eqnarray} \label{Wfact00}
W_i^{(0)} &=&
  h_i(p_X^-,p_X^+,m_b,\mu_i) \:
  \int_{0}^{p_X^+} \! \! dk^+\:
   {\cal J}^{(0)}(p^-\,, k^+,\mu_i )\:
   f^{(0)}(k^+  \!+\!\overline\Lambda\!-\! p_X^+,\mu_i) \,,
\end{eqnarray}
where RG evolution from the hard scale to the intermediate scale gives
\begin{align}
 h_i(p_X^-,p_X^+,\mu_i) = U_H(\mu_i,\mu_b)\: h_i(p_X^-,p_X^+,\mu_b)\,,
\end{align}
and the results at $\mu=\mu_b$ are determined from the traces in Eq.~(\ref{LOTr}):
\begin{eqnarray} \label{h123}
  h_1(p_X^-,p_X^+,\mu_b) &=&
   \frac{1}{2} \Big( \big|{\cal C}_{9}\big|^2 
     + \big|{\cal C}_{10a}\big|^2 \Big) +
  \frac{2\, 
    \mbox{Re}\big[ {\cal C}_{7}\, {\cal C}_{9}^{\,*} \big]
  }{(1\!-\!\bar y_H)}\
  +
  \frac{2\,\big| {\cal C}_{7} \big|^2 }{(1\!-\!\bar y_H)^2}\,, \\[5pt]
  h_2(p_X^-,p_X^+,\mu_b) &=& 
    \frac{2 \, (1\!-\!u_H)}{(\bar y_H \!-\! u_H)} 
   \Big( \big|{\cal C}_{9}\big|^2 \!+\! \big|{\cal C}_{10a}\big|^2 
    \!+\! \mbox{Re} \big[{\cal C}_{10a}\, {\cal C}_{10b}^{\,*} \big] \Big)
   \!+\! \frac{\big|{\cal C}_{10b}\big|^2}{2}
  \!-\! \frac{8 \big|{\cal C}_{7}\big|^2 }
  {(1\!-\!\bar y_H)(\bar y_H \!-\! u_H)}  \,,
  \nn \\[5pt] 
  h_3(p_X^-,p_X^+,\mu_b) &=&
    \frac{-4\,\mbox{Re}[ {\cal C}_{10a}\, {\cal C}_{7}^{\,*}]    }
    {m_B (1-\bar y_H)(\bar y_H - u_H)} -
  \frac{  2\,\mbox{Re}[{\cal C}_{10a}\, {\cal C}_{9}^{\,*}]
    }
    {m_B (\bar y_H - u_H)}
     \,.\nn
\end{eqnarray}
Here ${\cal C}_i={\cal C}_i(p_X^-,p_X^+,\mu_b,\mu_0,m_b)$, so these hard
coefficients also depend on $m_b$ and have residual $\mu_0$ scale dependence.
Explicit formulae are given in Eq.~(\ref{Cresult}).  For convenience we define
\begin{align} \label{defnF0}
  F^{(0)}(p^+_X,p^-_X) & = U_H(\mu_i,\mu_b) \int_{0}^{p_X^+} \! \! dk^+\:
   {\cal J}^{(0)}(p^-, k^+,\mu_i )\:
   f^{(0)}(k^+  \!+\!\overline\Lambda\!-\! p^+_X,\mu_i)  \nn\\
 &=   p_X^+\,U_H(\mu_i,\mu_b) \int_{0}^{1} \! \! dz\:
   {\cal J}^{(0)}(p^-, z\, p_X^+,\mu_i )\:
   f^{(0)}\big(\overline\Lambda\!-\! p^+_X(1\!-\!z),\mu_i \big)
\,.
\end{align}
where $p_X^-=p^- + \bar\Lambda$.
In terms of this function,
\begin{align}
 W_i^{(0)} = h_i(p^+_X,p^-_X,\mu_b) \: F^{(0)}(p^+_X,p^-_X) \,.
\end{align}
We find that to NLL order
\begin{align} \label{F0NLL}
  F^{(0)}(p^+_X,p^-_X) & = 
  U_H(\mu_i,\mu_b) f^{(0)}\big(\overline\Lambda\!-\! p^+_X,\mu_i
  \big) \\[3pt]
  &\hspace{-1cm}
 + U_H(\mu_i,\mu_b) \frac{\alpha_s(\mu_i)C_F}{4\pi} \bigg\{ 
  \Big( 2 \ln^2\frac{p^- p_X^+}{\mu_i^2}
  \!-\!  3 \ln \frac{p^- p_X^+}{\mu_i^2} + 7-\pi^2 \Big)
  f^{(0)}\big(\overline\Lambda\!-\! p^+_X,\mu_i \big) \nn\\
 & + \int_0^1\! \frac{dz}{z}\: \Big[  4 \ln\frac{z\,p^- p_X^+}{\mu_i^2} -3 \Big]
  { \Big[f^{(0)}\big({\overline \Lambda}\!-\!p_X^+ (1\!-\!z),\mu_i\big)
   -f^{(0)}\big(\overline\Lambda\!-\!p_X^+,\mu_i\big)\Big]}
  \bigg\} \,.\nn
\end{align}
Note that, until we include the $\alpha_s$ corrections from the jet function,
$F^{(0)}$ is independent of $p_X^-$, so that all of this dependence is in the
$h_i(p^+_X,p^-_X,\mu_b)$ functions.

Now, the triply differential decay rate in Eq.~(\ref{dGamma3u}) becomes
\begin{align} \label{triplefinal}
 \frac{1}{\Gamma_0} \frac{d^3\Gamma}{ dx_H\, d\overline y_H\, du_H}  
    &= {24m_B} (\overline y_H\!-\! u_H)
   \Big\{(1\!-\! u_H)(1\!-\! \overline y_H) h_1 + 
    \frac12 (1\!-\!x_H\!-\!u_H)(x_H\!+\!\overline y_H\!-\!1) h_2 \nn\\
 & \hspace{0.5cm}
  + \frac{m_B}{2}\,(1\!-\! u_H)(1\!-\! \overline y_H)
     \big(2x_H\!+\! u_H \!+\!\overline y_H\!-\! 2\big) h_3 \Big\}
    F^{(0)}(m_Bu_H, m_B \overline y_H)
    \,,
\end{align}
with $h_{1,2,3}$ from Eq.~(\ref{h123}).
As a check on this result, one can make the substitutions 
\begin{eqnarray} \label{eq:check}
{\cal C}_{9a}  =  - {\cal C}_{10a} & = & 1/2\,,
\;\; {\cal C}_{7}   = {\cal C}_{10b} = 0 \,,
 \\
\frac{G_F \alpha}{\sqrt{2}\pi} \, V_{tb} V_{ts}^* & \rightarrow &
 \frac{4 G_F}{\sqrt{2}} V_{ub} \,,\nn
\end{eqnarray}
after which the $h_1$ and $h_2$ terms in Eq.~(\ref{triplefinal}) agree with
terms in the leading-order shape-function spectrum for $B\to
X_u\ell\bar\nu$~\cite{bm04,blw04}. The $h_3$ term for $B\to X_s\ell\ell$ was the
difference of products of left- and right-handed currents and so should not agree
in this limit.

\subsection{ $d^2\Gamma/dq^2dm_X^2$ Spectrum with $q^2$ and $m_X$ Cuts}

Next we discuss doubly differential rates and forward-backward 
asymmetries.  For $d^2\Gamma/dq^2 dm_X^2$ the rate is obtained from
Eq.~(\ref{triplefinal}) by integrating over $x_H$ and changing variables. In
terms of dimensionless variables $y_H=q^2/m_B^2$ and $s_H=m_X^2/m_B^2$ we have
\begin{eqnarray} \label{dGdysLO}
 \frac{1}{\Gamma_0}\ \frac{d^2\Gamma}{dy_H ds_H}
   & = & 
     H^{ys}(y_H,s_H) \: m_B F^{(0)} \Big(m_B u_H(y_H,s_H), 
         m_B \overline y_H(y_H,s_H) \Big)
    \,, \\
  \frac{1}{\Gamma_0}\ \frac{d^2A_{\rm FB}}{dy_H ds_H} 
  & = & K^{ys}(y_H,s_H)\:  m_B F^{(0)} \Big(m_B u_H(y_H,s_H), 
         m_B \overline y_H(y_H,s_H) \Big)
    \,, \nn
\end{eqnarray}
where 
\begin{align}
 H^{ys}(y_H,s_H) &= {2} \sqrt{(1\!-\!y_H\!+\! s_H)^2\!-\! 4\, s_H}\:
   \Big\{ 12 y_H h_1  
   +\big[(1\!-\!y_H\!+\! s_H)^2 \!-\! 4 s_H\big] h_2 \Big\} \,,\\
 K^{ys}(y_H,s_H) &= 
     6  y_H \big[ (1\!-\! y_H \!+\! s_H)^2 -4 s_H \big] h_3
    \nn
\end{align}
and we need to substitute $h_{1,2,3}$ from Eq.~(\ref{h123}) and $u_H(y_H,s_H)$
and $\overline y_H(y_H,s_H)$, as given in Eq.~(\ref{eq:yu}). When one takes 
experimental cuts on $q^2$ and $m_X^2$,
\begin{align}
  y_H^{\rm min} < y_H < y_H^{\rm max} \,,\qquad
  0 < s_H < s_H^0 \,,
\end{align} 
the limits on the doubly differential rate and forward-backward asymmetry in
Eq.~(\ref{dGdysLO}) are
\begin{align}
   & 1) \qquad\ \ \  y_H^{\rm min} \le y_H \le y_H^{\rm max} \,,
       &0 \le &\: s_H \le {\rm min} \big\{ s_H^0 \,, (1\!-\!\sqrt{y_H})^2 \big\} 
     \,,\nn\\
   & 2) \qquad \ \ \   0 \le s_H \le s_H^0 \,,\qquad
     & y_H^{\rm min} \le &\: y_H \le {\rm min} 
      \big\{ y_H^{\rm max} \,, (1\!-\!\sqrt{s_H})^2\big\}
       \,,
\end{align}
depending on the desired order of integration.

\subsection{ $d^2\Gamma/dm_X^2dp_X^+$ Spectrum with $q^2$ and $m_X$ Cuts}

The hadronic invariant-mass spectrum and forward-backward asymmetry
can be obtained by integrating the doubly differential spectra
\begin{eqnarray} \label{dGdsLO}
\frac{1}{\Gamma_0}\ \frac{d^2\Gamma}{ds_H du_H}
   & = &  H^s(s_H,u_H)  \: m_B F^{(0)}\Big(m_B u_H , m_B\frac{s_H}{u_H} \Big)
    \,, \nn\\
\frac{1}{\Gamma_0}\ \frac{d^2 A_{\rm FB} }{ds_H du_H}
   & = &  K^s(s_H,u_H)  \: m_B F^{(0)}\Big(m_B u_H , m_B\frac{s_H}{u_H} \Big)
\end{eqnarray}
over $u_H$. Here
\begin{align} \label{dGdsLO2}
H^s(s_H,u_H) & =  
 \frac{4 (s_H-u_H^2)^2}{(u_H-s_H)u_H^4} 
  \bigg\{
  (1\!-\!u_H)(u_H \!-\! s_H)  (3 u_H \!-\! 2 s_H \!-\! u_H^2)
  \big( | {\cal C}_{9}|^2+ | {\cal C}_{10a}|^2  \big) 
 \\[3pt]
& 
  \quad
 + 4\, u_H (3 u_H \!-\! s_H \!-\!  2u_H^2)| {\cal C}_{7}|^2
 + 12 u_H (1\!-\! u_H)(u_H\!-\!s_H)  
     \mbox{Re}\big[{\cal C}_{7}\, {\cal C}_{9}^{\,*}\big] 
 \nn \\[3pt] 
& 
  \quad
  + (1\!-\! u_H)(u_H\!-\!s_H)(s_H\!-\!u_H^2) 
       \mbox{Re}\big[{\cal C}_{10a}\, {\cal C}_{10b}^{\,*}\big] \,
  + \frac{(u_H\!-\!s_H) (s_H\!-\!u_H^2)^2 }{4 u_H} \, | {\cal C}_{10b}|^2
   \bigg\}\nn \,, \nn \\[5pt]
 K^{s}(s_H,u_H) &=   
  \frac{-12(s_H\!-\! u_H^2)^2 (u_H\!-\!s_H) (1\!-\!u_H)}{u_H^4}\: 
   \bigg\{ 
    {\rm Re}\big[  {\cal C}_9\, {\cal C}_{10a}^* \big] +
   \frac{2 u_H \, }{u_H\!-\! s_H}\:  {\rm Re}\big[  {\cal C}_7\, {\cal C}_{10a}^* \big]
   \bigg\} \,,
    \nn
\end{align}
and the limits with $q^2$ and $m_X$ cuts are
\begin{align}
 & 0 \le s_H \le s_H^0\,,\qquad
   {\rm max}\Big\{ s_H\,, u_1(s_H) \Big\}
   \le u_H \le {\rm min}\: \Big\{ \sqrt{s_H} \,, 
     u_2(s_H)\Big\}\,,\nn\\
 & u_1(s_H) = \frac{1\!+\!s_H\!-\!y_H^{\rm min} \!-\!
   \sqrt{(1\!+\!s_H\!-\!y_H^{\rm min})^2\!-\!4 s_H}}{2} \,, \nn\\
 &  u_2(s_H) = \frac{1\!+\!s_H\!-\!y_H^{\rm max} \!-\!
   \sqrt{(1\!+\!s_H\!-\!y_H^{\rm max})^2\!-\!4 s_H}}{2}\,.
\end{align}

\subsection{ $d^2\Gamma/dq^2dp_X^+$ Spectrum with $q^2$ and $m_X$ Cuts}

From Eqs.~(\ref{dGamma2up}) and the above results, we can obtain the dilepton
invariant-mass spectrum and forward-backward asymmetry, for example by
integrating the doubly differential spectra
\begin{eqnarray} \label{dGdyLO}
\frac{1}{\Gamma_0}\ \frac{d^2\Gamma}{dy_H du_H}
   & = & 
     H^y(y_H,u_H) \: m_B F^{(0)} \Big(m_B u_H, m_B\,
         \frac{1\!-\!y_H\!-\! u_H}{1\!-\! u_H} \Big) \,, \\
\frac{1}{\Gamma_0}\ \frac{d^2 A_{\rm FB}}{dy_H du_H}
   & = & 
     K^y(y_H,u_H) \: m_B F^{(0)} \Big(m_B u_H, m_B\,
         \frac{1\!-\!y_H\!-\! u_H}{1\!-\! u_H} \Big) \, \nn 
\end{eqnarray}
over $u_H$. Here 
\begin{align} \label{HyKy}
H^y(y_H,u_H) & = \frac{4[(1\!-\!u_H)^2\!-\! y_H]^2}{y_H(1\!-\! u_H)^3}
  \bigg\{
  y_H [(1\!-\!u_H)^2 \!+\! 2 y_H]\,\big( |{\cal C}_{9}|^2+|{\cal
   C}_{10a}|^2\big)
  \\[3pt]
 &   \qquad\qquad
 + \,[8(1\!-\!u_H)^2\!+\! 4y_H]\, |{\cal C}_{7}|^2
 +12 y_H (1\!-\!u_H)\mbox{Re}\big[{\cal C}_{7}\, {\cal C}_{9}^{\,*} \big] 
  \nn\\[3pt]
 &\qquad\qquad + \, y_H [(1\!-\!u_H)^2 \!-\!  y_H]\,\mbox{Re}\big[ {\cal
   C}_{10a} {\cal C}_{10b}^* \big]
   + \, \frac{y_H [(1\!-\!u_H)^2 \!-\!  y_H]^2}{4(1\! -\! u_H)^2}\: \big|
   {\cal C}_{10b} \big|^2 \,
  \bigg\}\,, \nn\\[4pt]
 K^{y}(y_H,u_H) &= 
    \frac{-12 y_H\,[(1\! -\! u_H)^2\!-\! y_H]^2 }{(1\! -\! u_H)^3}\: \bigg\{
    {\rm Re}\big[ {\cal C}_9\, {\cal C}_{10a}^* \big]
   + \frac{2(1\!-\! u_H)    }{y_H}\: {\rm Re}\big[  {\cal C}_7\, {\cal
     C}_{10a}^* \big]   
   \bigg\}\,,
    \nn
\end{align}
and the limits of integration with cuts are
\begin{align}
  y_H^{\rm min} < y_H < y_H^{\rm max} \,,\quad 
   0 \le u_H \le {\rm min}\: \bigg\{ 1\!-\!\sqrt{y_H} \,, 
    \frac{1\!+\!s_H^0\!-\!y_H \!-\!
   \sqrt{(1\!+\!s_H^0\!-\!y_H)^2\!-\!4 s_H^0}}{2}\, \bigg\}.
\end{align}
The opposite order of integration is also useful:
\begin{align} \label{uHlimits}
 &  0 \le u_H \le 1 \,,\quad
   y_1(u_H) < y_H < y_2(u_H)\,,\nn\\
 & y_1(u_H) = {\rm max} \bigg\{ y_H^{\rm min}, 
  \frac{(1\!-\!u_H)(u_H\!-\! s_H^0)}{u_H}  \bigg\} \,,\quad
   y_2(u_H) = {\rm min} \Big\{ y_H^{\rm max}, (1\!-\! u_H)^2
   \Big\} \,.
\end{align}

The doubly differential rate can also be expressed in terms of the coefficients
$C_9^{\rm mix}$, $C_7^{\rm mix}$, and $C_{10}$. This is one step closer to the
short-distance coefficients $C_9$, $C_7$, and $C_{10}$ of $H_W$, which we wish
to measure in order to test the Standard Model predictions for the
corresponding FCNC 
interactions. Substituting Eq.~(\ref{Cresult}) into Eq.~(\ref{HyKy}) gives
\begin{align}
H^y( y_H, u_H) & =
   \frac{4\big[(1\!-\! u_H)^2\!-\! y_H\big]^2}{(1\!-\! u_H)^3}
  \bigg\{  \big| C_7^{\rm mix}(s,\mu_0)\big|^2 \Big[ 4 \Omega_C^{\,2}(s,\mu_b)
   \!+\! \frac{8(1\! -\! u_H)^2}{y_H}\, \Omega_D^{\,2}(s,\mu_b) \Big]
  \nn \\
 & 
  + \big[ \big|C_9^{\rm mix}(s,\mu_0)\big|^2 \!+\! C_{10}^{\,2} \big] 
  \Big[ 2 y_H\, \Omega_A^{\,2}(s,\mu_b) \!+\! (1\! - \! u_H)^2\,
  \Omega_B^{\,2}(y_H,u_H,s,\mu_b) \Big] \nn\\
 & 
   + {\rm Re}\big[ C_7^{\rm mix}(s,\mu_0) C_9^{\rm mix}(s,\mu_0)^*\big]
   \Big[ 12(1\! -\! u_H) \Omega_E(s,\mu_b) \Big]
 \bigg\}
 \nn\\[4pt]
 K^y( y_H, u_H) &= 
   \frac{ -12 y_H\,[(1\! -\! u_H)^2\!-\! y_H]^2 }{(1\! -\! u_H)^3}\: \bigg\{
     {\rm Re}\big[  {C}_9^{\rm mix}(s,\mu_0)\, {C}_{10}^* \big]
     \Omega_A^2(s,\mu_b) \nn\\
 &   +
   \frac{2(1\!-\! u_H)}{y_H}\: 
    {\rm Re}\big[ C_7^{\rm mix}(s,\mu_0)\, {C}_{10}^* \big]
   \Omega_A(s,\mu_b) \Omega_D(s,\mu_b)  
   \bigg\}\,,
\end{align}
where $s=q^2/m_b^2$ and
\begin{align}
  \Omega_A &= 1 \!+\! \frac{\alpha_s(\mu_b)}{\pi}\, \omega_a^V(s,\mu_b) 
   \,,\\
  \Omega_B &= 1 \!+\! \frac{\alpha_s(\mu_b)}{\pi} \Big[
    \omega_a^V(s,\mu_b) \!+\! \omega_c^V(s,\mu_b) \!+\! 
   \frac{(1\!-\! u_H)^2\!-\! y_H}{2(1\!-\! u_H)^2}
   \omega_b^V(s,\mu_b)
   \Big] \,,
  \nn\\ 
   \Omega_C &= 1 \!+\! \frac{\alpha_s(\mu_b)}{\pi}\, \Big[
    \omega_a^T(s,\mu_b) - \omega_b^T(s,\mu_b) - \omega_d^T(s,\mu_b) \Big]
   \,, \nn \\
  \Omega_D &= 1 \!+\! \frac{\alpha_s(\mu_b)}{\pi} \Big[ 
    \omega_a^T(s,\mu_b) - \omega_c^T(s,\mu_b) 
   -\frac{(1\!-\! u_H)^2\!+\! y_H}{2(1\!-\! u_H)^2}
   \omega_b^T(s,\mu_b) \Big] \,,
  \nn\\
  \Omega_E &= \big( 2 \Omega_A\Omega_D + \Omega_B \Omega_C\big)/3 \,. \nn
\end{align}
This is the form that turned out to be  the most useful for the analysis in
Ref.~\cite{llst}.

\subsection{Numerical Analysis of Wilson Coefficients} \label{numbers}

\begin{figure}[t!]
  \vskip .2cm \centerline{ 
   \mbox{\epsfysize=8.truecm \hbox{\epsfbox{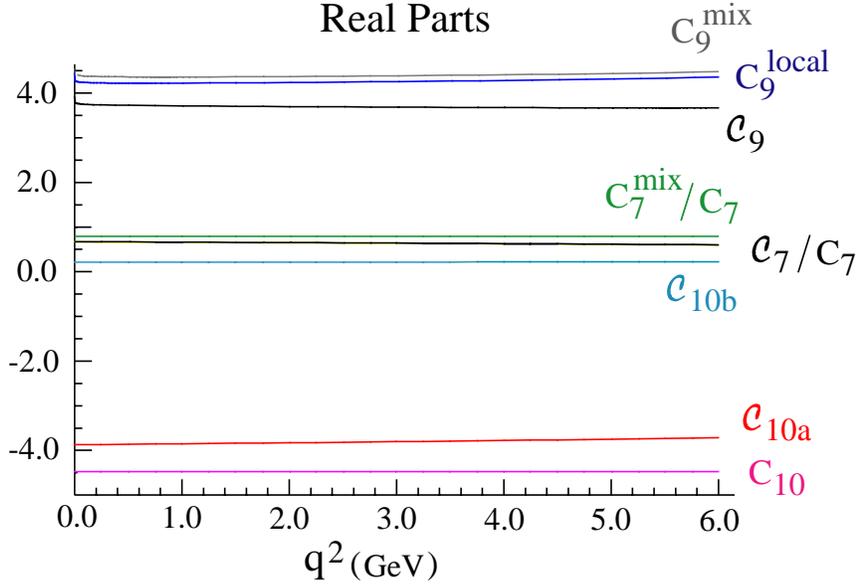}} }\quad
   } 
  \vskip -.4cm 
  {\caption[1]{Comparison of the real part of Wilson coefficients at
      $\mu_0=\mu_b=4.8\,{\rm GeV}$ with $m_c/m_b=0.292$, $\overline {m_b}(\mu_0)
      =4.17\,{\rm GeV}$, and $m_b=4.8\,{\rm GeV}$. For ${\cal C}_9$, ${\cal
        C}_7$, and ${\cal C}_{10b}$ we take $p_X^+=0$.}
\label{fig_CC} }
 \vskip -0.1cm
\end{figure}

As shown in Fig.~\ref{fig_kinematics}, for the small-$q^2$ window ($q^2< 6\,{\rm
  GeV^2}$) we have $p_X^+\ll p_X^-$. Generically, the hard contributions in
${\cal C}_9$, ${\cal C}_7$, and ${\cal C}_{10a,10b}$ from our split-matching
procedure depend on the variable $q^2$. In Fig.~\ref{fig_CC} we plot the $q^2$
dependence of the real part of the coefficients and see that there is in fact
very little numerical change over the low-$q^2$ window. Here 
${\rm Re}\big[ C_9^{\rm local} \big]$ varies by $\pm 1.5\%$, 
${\rm Re}\big[ C_9^{\rm mix} \big]$ by $\pm 1 \%$, and the real parts of
$\{{\cal C}_9, {\cal C}_7, {\cal C}_{10a}, {\cal C}_{10b}\}$ by 
$\{\pm 1\%, \pm 5\%, \pm 2\%, \pm 3\%\}$. 
The imaginary parts are either very small or also change by only a few
percent over the low-$q^2$ window. The analytic formulae for the
$q^2$ dependence mean that there is no problem keeping the exact dependence, but
this does make it necessary to perform integrals over regions in $q^2$
numerically. A reasonable first approximation can actually be obtained by fixing
a constant $q^2$ in the hard coefficients, while keeping the full $q^2$
dependence elsewhere.

Since the coefficients change very little with $q^2$ we continue our numerical
analysis by fixing $q^2 =3\,{\rm GeV}^2$. If we then take
$\mu_0=\mu_b=m_b=4.8\,{\rm GeV}$, $\overline {m_b}(\mu_0)=4.17\,{\rm GeV}$,
$m_c/m_b=0.292$ and $p_X^+=0$ we find that Eq.~(\ref{Cresult}) gives
\begin{align}
 {\cal C}_9 &=0.826\: C_9^{\rm mix}
            +0.097 \: C_7^{\rm mix} 
    = 3.448\: \frac{C_9^{\rm mix}}{C_9^{\rm NDR}} 
            -0.030\: \frac{C_7^{\rm mix}}{C_7^{\rm NDR}} \,,\nn\\
 {\cal C}_7 &= 0.823\: C_7^{\rm mix}
            +0.001 \: C_9^{\rm mix} 
  =  -0.239\: \frac{C_7^{\rm mix}}{C_7^{\rm NDR}} 
           +0.005\: \frac{C_9^{\rm mix}}{C_9^{\rm NDR}} \,.
\end{align}
These numbers indicate that, despite the entanglement of $C_{7,9}^{\rm mix}$ in
${\cal C}_{7,9}$ due to $\alpha_s(m_b)$ corrections, numerically ${\cal C}_9$ is 
dominated by $C_9$ and ${\cal C}_7$ is dominated by $C_7$ in the Standard Model.

For the coefficients at $q^2=3.0\,{\rm GeV}^2$, with the other parameters as above,
we have
\begin{align} \label{Cnums}
  {C}_9^{\,\rm mix} &= 4.487 + 0.046 i \,,
  & C_{7}^{\,\rm mix} & = -0.248 \,,
  \nn\\
  {\cal C}_9(u_H=0) &= 3.683+0.038 i \,,
   &{\cal C}_7(u_H=0) &= -0.198 + 6\times 10^{-5}i\,, 
  \nn\\
  {\cal C}_9(u_H=0.2) &= 3.663+0.038 i 
   & {\cal C}_7(u_H=0.2) &= -0.193 +   10^{-4}i\,, 
  \nn\\
  {\cal C}_{10a} &= -3.809\,,
  & {\cal C}_{10b}(u_H=0) &= 0.214\,,\nn\\
   &\phantom{=} & {\cal C}_{10b}(u_H=0.2) &= 0.237\,.
\end{align}
The relevant range of $p_X^+$ in Fig.~\ref{fig_kinematics} gives $0\le u_H \le
0.2$.  From the above numbers it is easy to see that the $u_H$ dependence of
${\cal C}_9$, ${\cal C}_7$, and $C_{10b}$ is very mild over the range of
interest. The perturbative $\alpha_s$ corrections due to $\omega_i^{V,T}$ reduce
both ${\cal C}_9$ and ${\cal C}_7$ by 17\% relative to $C_9^{\rm mix}$ and
$C_7^{\rm mix}$ respectively, and ${\cal C}_{10a}$ by 15\%. This can be seen
both in Fig.~\ref{fig_CC} and in Eq.~(\ref{Cnums}), when one notes that
$C_{10}=-4.480$.  Comparing with coefficients in the local OPE, we note that the
$\omega_{\rm semi}^{\rm OPE}$ factor, which accounts for the difference between
$C_9^{\rm local}$ and $C_9^{\rm mix}$, is significantly smaller than the
combination of $\alpha_s$ corrections in the $\omega_i^V$ terms that shifts
${\cal C}_9$ from its lowest-order value.

In quoting the above numbers, we have not varied the scales $\mu_0$ and $\mu_b$.
The main point was to compare the size of the hard corrections in the shape
function and local OPE regions, and to see how much deviation from $C_{7,9}^{\rm
  mix}$ they cause. The dependence on $\mu_0$ for the ${\cal C}_i$ is similar to
that in the local OPE analysis at NLL~\cite{misiak93,bm95} and will be reduced
by a similar amount when the full NNLL expressions are included in $C_{7,9}^{\rm
  mix}$.  The $\mu_b$ dependence of the ${\cal C}_i$ is fairly strong because of
the appearance of double logarithms, but it is canceled by the $\mu_b$
dependence in the function $F^{(0)}$, which contains the NLL jet and shape
functions.


\section{Conclusion  } \label{conclusion}

In this paper we have performed a model-independent analysis of $B\to
X_s\ell^+\ell^-$ decays with cuts giving the small-$q^2$ window and an $m_X$ cut
to remove $b\to c$ backgrounds. These cuts put us in the shape function region.
We analyzed the rate for the formal counting with $q^2\sim \lambda^0$ and $m_X^2
\sim \lambda^2$ and showed that the same universal shape function as in $B\to
X_u\ell\bar\nu$ and $B\to X_s\gamma$ is the only non-perturbative input needed
for these decays. We also developed a new effective-theory technique of split
matching. Split matching between two effective theories is done not at a single
scale $\mu$, but rather at two nearby scales. For $B\to X_s\ell^+\ell^-$ this
allowed us to decouple the perturbation-theory analysis above and below $m_b$,
which simplifies the organization of the $\alpha_s$ contributions.

In Section~\ref{results} we presented the leading-power triply differential
spectrum and doubly differential forward-backward asymmetry with
renormalization-group evolution and matching to ${\cal O}(\alpha_s)$. Above the
scale $m_b$, we restricted our analysis to include the standard NLL terms from
the local OPE, but illustrated how terms from NNLL can be incorporated. Below
$m_b$ we considered running to NLL and matching at one-loop (NNLL evolution will
be straightforward to incorporate if desired). We then computed several
phenomenologically relevant doubly differential spectra with phase-space cuts on
$q^2$ and $m_X$ (from which the singly differential spectra can be obtained by
numerical integration).  In section~\ref{results}E we discussed the numerical
size of our perturbative hard coefficients and compared them to the local OPE
results.

Our results for the doubly differential rate in Eqs.~(\ref{dGdyLO}) and
(\ref{HyKy}), together with $F^{(0)}$ from Eq.~(\ref{F0NLL}), determine the 
shape-function-dependent rate for $B\to X_s\ell^+\ell^-$.  Using as input a 
result for the non-perturbative shape function $f^{(0)}$ from a fit to the 
$B\to X_s\gamma$ spectrum or from $B\to X_u\ell\bar\nu$ gives a 
model-independent result for $B\to X_s\ell^+\ell^-$ with phase-space cuts.  A 
full investigation of the $m_X$-cut dependence and phenomenology is carried 
out in a companion publication~\cite{llst}. An intriguing universality of the 
cut dependence is found, which makes the experimental extraction of 
short-distance Wilson coefficients in the presence of cuts much simpler.  An 
extension of the analysis of this paper to include subleading shape-function 
effects will be presented in the near future~\cite{lst2}.

\acknowledgments We thank Z.~Ligeti and F.~Tackmann for helpful discussions and
for collaboration on a related analysis of $B\to X_s \ell\ell$, and D.~Pirjol
for comments on the manuscript.  This work was
supported in part by the U.S.  Department of Energy (DOE) under the cooperative
research agreement DF-FC02-94ER40818 and by the Office of Nuclear Science.  I.S.
was also supported in part by the DOE Outstanding Junior Investigator program
and the Sloan foundation.

\appendix

\section{Wilson Coefficients} \label{app:defs}

The coefficients and functions that appear in Eq.~(\ref{eq:C7910}) 
are defined as follows \cite{bm95}.
\begin{eqnarray}
C_7(M_W) & = & - \frac{1}{2} A(m_t^2/M_W^2) \,, \nn \\
C_8(M_W) & = & - \frac{1}{2} F(m_t^2/M_W^2) \,, \nn \\
Y(x) & = & C(x) - B(x) \,, \nn \\
Z(x) & = & C(x) + \frac{1}{4} D(x) \,, \nn \\
A(x) & = & \frac{x(8x^2+5x-7)}{12(x-1)^3} 
 + \frac{x^2(2-3x)}{2(x-1)^4} \ln x \,, \nn \\
B(x) & = & \frac{x}{4(1-x)} + \frac{x}{4(x-1)^2} \ln x \,, \nn \\
C(x) & = & \frac{x(x-6)}{8(x-1)} + \frac{x(3x+2)}{8(x-1)^2} \ln x \,, \nn \\
D(x) & = & \frac{-19 x^3 + 25 x^2}{36 (x-1)^3}
+ \frac{x^2 (5 x^2 - 2 x - 6)}{18 (x-1)^4} \ln x- \frac{4}{9} \ln x \,, \nn \\
E(x) & = & \frac{x (18 -11 x - x^2)}{12 (1-x)^3} 
 + \frac{x^2 (15 - 16 x + 4 x^2)}{6 (1-x)^4} \ln x-\frac{2}{3} \ln x \,, \nn \\
F(x) & = & \frac{x(x^2-5x-2)}{4(x-1)^3} + \frac{3x^2}{2(x-1)^4} \ln x \,, \nn 
\end{eqnarray}
and 
\begin{equation*}
\begin{array}{rrrrrrrrrl}
t_i = (\!\! & 2.2996, & - 1.0880, & - \frac{3}{7}, & -
\frac{1}{14}, & -0.6494, & -0.0380, & -0.0186, & -0.0057 & \!\!) \,,
\vspace{0.1cm} \\

a_i = (\!\! & \frac{14}{23}, & \frac{16}{23}, & \frac{6}{23}, 
& - \frac{12}{23}, & 0.4086, & -0.4230, & -0.8994, & 0.1456 & \!\!) \,, 
\vspace{0.1cm} \\

p_i = (\!\! & 0, & 0, & -\frac{80}{203}, &  \frac{8}{33}, &
0.0433, &  0.1384, & 0.1648, & - 0.0073 & \!\!) \,, \vspace{0.1cm} \\
                                                                                
\rho^{\rm NDR}_{i} = (\!\! & 0, & 0, & 0.8966, & - 0.1960, &
- 0.2011, & 0.1328, & - 0.0292, & - 0.1858 & \!\!) \,, \vspace{0.1cm} \\
                                                                                
s_i = (\!\! & 0, & 0, & - 0.2009, &  -0.3579, &
0.0490, & - 0.3616, & -0.3554, & 0.0072 & \!\!)  \,, \vspace{0.1cm} \\
                                                                                
q_i = (\!\! & 0, & 0, & 0, &  0, &
0.0318, & 0.0918, & - 0.2700, & 0.0059 & \!\!) \,.
\end{array}
\end{equation*}

\section{The Case of Collinear $q^2$} \label{App:q2jet}

In the body of the paper we used $q^2\sim \lambda^0$. We were free to choose
this counting since the power counting for the leptonic variable $q^2$ does not
affect the counting for $p_X^\pm$ in the shape function region.  (The only
restriction was not to have $q^2$ too close to $m_b^2$.) However, we are free to
consider other choices. In this appendix we consider how our analysis will
change if we instead take $q^2\sim \lambda^2$. With this scaling, new physical
degrees of freedom are needed at leading order in SCET, making the analysis more
complicated. In particular we must consider graphs with quark fields that are
collinear to the collinear photon (or dilepton pair), since with this power
counting we have $(q^0)^2 \gg q^2$.

\begin{figure}[t!]
 \vskip .2cm
 \centerline{
  \mbox{\epsfysize=4.truecm \hbox{\epsfbox{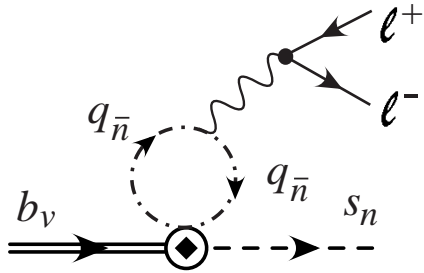}} }
  }
 \vskip -.4cm
 {\caption[1]{Additional graphs in SCET for the matching computation for the 
case where $q^2\sim \lambda^2$.}
\label{fig_scetm2} }
 \vskip -0.1cm
\end{figure}

An example of a new {\em nonzero} graph is the one generated by four-quark
operators within SCET, as shown in Fig.~\ref{fig_scetm2}, which involve these
additional degrees of freedom.  In this graph we have a light-quark loop of
collinear-${\bn}$ fields that are collinear to the virtual photon. The
presence of this type of diagram changes the hard matching at $\mu_b=m_b$. It
also means that we have a more complicated pattern of operator mixing within
SCET, since divergences in the displayed diagram will cause an evolution for
$C_9$, etc. Therefore, the running below $m_b$ will no longer be universal. In
the presence of these diagrams the jet function will also no longer be given by
a single bilinear operator, since it will also involve some contributions
with a factorized matrix element of $\bn$-fields, which are also integrated out
at $p^2\sim m_b\Lambda_{\rm QCD}$.  Finally, the appearance of these additional
degrees of freedom might also affect the number of non-perturbative shape
functions that appear in the factorization theorem.  It would be interesting to
carry out a detailed analysis of this $q^2\sim \lambda^2$ case in the future.

In $B\to X_s\gamma$ at lowest order, the analog of the graph in
Fig.~\ref{fig_scetm2} vanishes at one-loop order, and this argument can be
extended to include higher orders in $\alpha_s$~\cite{neubert}. This relies on
the fact that here $q^2=0$ and does not generate a scale. We find that the same
reasoning does not apply for $B\to X_s\ell\ell$ for parametrically small but
finite $q^2$.

Finally, we comment on the possibility of penguin charm-loop effects. In our
analysis we integrated out the charm loops at the same time as the bottom loops.
This is reasonable when treating $q^2\sim \lambda^0$.  One could also consider
the case $m_c^2 \sim m_b\Lambda$, which is also reasonable numerically. This
type of power counting was considered for the simpler case of $B\to
X_c\ell\bar\nu$ decays with energetic $X_c$ in Ref.~\cite{Boos:2005by} and it
would be interesting to extend this to $B\to X_s\ell\ell$. We remark that the
problematic region for $B\to \pi\pi$ factorization
theorems~\cite{Bauer:2004tj,Feldmann:2004mg,Beneke:2004bn,Bauer:2005wb}, which
is near the charm threshold, $q^2\sim 4m_c^2$, is not relevant for our analysis.
The experimental cuts on $q^2$ explicitly remove the known large contributions
from this region.

\end{document}